\documentclass[english,prb,twocolumn,aps]{revtex4-1}
\usepackage[T1]{fontenc}
\usepackage[latin9]{inputenc}
\usepackage{float}
\usepackage{graphicx}
\usepackage{amssymb}

\makeatletter

\providecommand{\tabularnewline}{\\}

\usepackage{float}

\makeatletter

\usepackage{array}

\usepackage{subfigure}

\usepackage{float}

\usepackage{textcomp}

\usepackage{esint}



\makeatother

\usepackage{babel}
\makeatother

\begin{document}

\title{Theoretical Study of Nuclear Spin Polarization and Depolarization
in Self-Assembled Quantum Dots }

\author{Chia-Wei Huang and Xuedong Hu}

\affiliation{Department of Physics, University at Buffalo, The State University
of New York, Buffalo, NY 14260-1500, USA}

\begin{abstract}
We investigate how the strain-induced nuclear quadrupole interaction influences
the degree of nuclear spin polarization in self-assembled quantum
dots. Our calculation shows that the achievable nuclear spin polarization
in In$_{x}$Ga$_{1-x}$As quantum dots is related to the concentration
of indium and the resulting strain distribution in the dots. The interplay
between the nuclear quadrupole interaction and Zeeman splitting leads
to interesting features in the magnetic field dependence of the nuclear
spin polarization. Our results are in qualitative agreement with measured
nuclear spin polarization by various experimental groups.
\end{abstract}
\maketitle

\section{Introduction}

Nuclear spin dynamics has been studied extensively in many fields.\cite{Slichter1992,Abragam1961}
In recent years, nuclear spin dynamics in semiconductor quantum dots
has attracted intense interest because of the excellent quantum coherence
properties of nuclear spins. Indeed, nuclear spins in nanostructures
have been suggested as qubits for a quantum computer\cite{Kane1998}
and for use as quantum memory.\cite{Taylor2003} For either purpose, high
degree of nuclear spin polarization (NSP) is a pre-requisite.

Nuclear spins in nanostructures also form an important environment
for electron spins, which have been proposed as a candidate for qubits.\cite{Loss1998}
Through the hyperfine interaction the nuclear spins create a spatially
and temporally fluctuating magnetic field for the electron spins,
which leads to spin decoherence.\cite{Merkulov2002,Witzel2005,Coish2004,Yao2006,Erlingsson2004,Khaetskii2002}
It has been suggested theoretically that such decoherence could potentially
be suppressed if nuclear spin fluctuations are suppressed,\cite{Burkard1999,Klauser2006,Ramon2007}
and one way to realize such suppression is via dynamic nuclear spin
polarization. Furthermore, the coupled electron-nuclear spin problem
is an intriguing example of a quantum many-body problem, and is still
not solved completely.

Dynamic nuclear spin polarization (DNSP) has been studied for many
decades.\cite{Meier1984} It has been demonstrated in semiconductor
quantum wells\cite{Dobers1988,Kane1992,Smet2002} and quantum dots
\cite{Gammon1996,Ono2004,Baugh2007,Lai2006,Petta2008,Danon2009}
through a variety of experiments. Physically, DNSP can be achieved
either electrically or optically, where a pumped electron can transfer
its spin polarization to nuclear spins via the contact hyperfine interaction.
A range of values for nuclear spin polarization has been reported by several experimental groups. For
example, using electrically controlled DNSP, Petta \textit{et al.}
reported approximately 1 \% NSP in lateral coupled GaAs double quantum
dots;\cite{Petta2008} while Baugh \textit{et al.} reported 40 \%
NSP in vertical coupled GaAs quantum dots with 5\% In.\cite{Baugh2007}
With optically pumped DNSP, Gammon \textit{et al.} reported 60
\% NSP in interface fluctuation GaAs quantum dots,\cite{Gammon1996}
while recent experiments have achieved NSP in In$_{x}$Ga$_{1-x}$As
self-assembled quantum dots at various magnetic fields. In particular,
approximately $10\sim20\%$ of NSP is created in In$_{0.9}$Ga$_{0.1}$As
quantum dots < 1 T,\cite{Lai2006} 40\% in In$_{0.6}$Ga$_{0.4}$As
at around 2 T,\cite{Braun2006,Eble2006,Tartakovskii2007} and 80\%
in In$_{0.9}$Ga$_{0.1}$As at 5 T. \cite{Maletinsky2008} It is
evident that the experimental results vary greatly as experimental
conditions and physical systems are varied. So far there has been
no systematic theoretical studies of NSP and how it depends on the
various system parameters such as applied field and material composition.

In this paper we study dynamic nuclear spin polarization in In$_{x}$Ga$_{1-x}$As
quantum dots via optical pumping of confined electrons. These self-assembled
dots are generally highly strained, and we are particularly interested
in the NSP of these dots in different strain environments. Specifically,
the strain breaks the cubic symmetry of the crystal lattice and creates an electric field
gradient which couples to the nuclear quadrupole moment,\cite{Dzhioev2007}
which in turn leads to mixing of nuclear spin eigenstates. We use
a simplified model of the quantum dot where the electric field gradient
is axially symmetric. We first study how the As NSP responds to
various strain strengths, angles and cotunneling constants in magnetic
fields. For the electron-nuclear spin transfer, we consider both phonon-assisted
and cotunneling-assisted spin flip processes. Lastly, we consider
NSP of In$_{x}$Ga$_{1-x}$As quantum dots with different compositions.

The paper is organized as follows. We describe the scheme of DNSP
in section \ref{sec:Scheme-of-Nuclear}, and our model Hamiltonian
in section \ref{sec:Model}. We show our results of nuclear spin polarization
in As nuclei and in different compositions of In$_{x}$Ga$_{1-x}$As
quantum dots at various magnetic fields in section \ref{sec:Calculation}.
We discuss some interesting features related to our calculation in
section \ref{sec:Discussion}. Finally, we summarize our results in
section \ref{sec:Conclusion} and draw our conclusions.

\section{\label{sec:Scheme-of-Nuclear}Scheme of Nuclear spin polarization}

\begin{figure}[t]

\begin{raggedleft}
\includegraphics[scale=0.5]{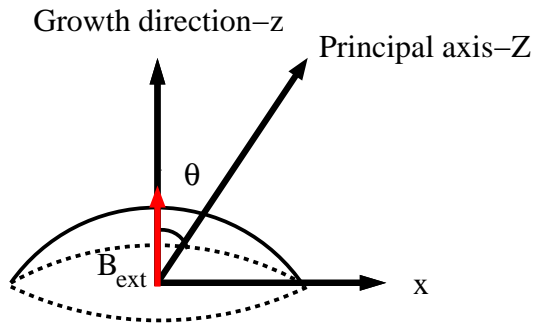}
\par\end{raggedleft}

\caption{\label{fig:sadthe-geometry-of}A model of a self-assembled quantum dot. We assume the field gradients in the self-assembled quantum dot
are axially symmetric. The largest electric field gradient component ${V}_{ZZ}$ is along the principal axis $Z$, which deviates from the
growth direction (the $z$-axis) by an angle $\theta$.  The external magnetic field is assumed to be along the $z$-axis. }

\end{figure}

In$_{x}$Ga$_{1-x}$As self-assembled quantum dots (SAQDs) are formed
by a strain-driven process, where the strain arises from the lattice
mismatch between the InAs deposition layers and the GaAs substrate.
The strain in the quantum dots (QDs) breaks the lattice symmetry and
creates electric field gradients in the dots. The shape of In$_{x}$Ga$_{1-x}$As
QDs varies among experiments, ranging from pancake-like to pyramid-like
and dome-like. The resulting distribution of electric field gradients
thus also differs from dot to dot. Even in the same dot, the strain
distribution is not uniform. For example, the strain at the edge of
a quantum dot is generally larger than at the center of the dot.\cite{Williamson1999}
Therefore nuclear spins in different regions of a QD experience electric
field gradients of different strengths and directions. For an estimate
of NSP, we start with a simplified model of a pancake-like cylindrically
symmetric QD as shown in Fig.~\ref{fig:sadthe-geometry-of}. The electric field
gradients in such a dot are thus axially symmetric. The largest electric
field gradient ${V}_{ZZ}$ is along the principal axis $Z$, which is defined
to be normal to the pancake surface. For instance,
for a lattice site in the $xz$ plane, $Z$ would be in the $xz$
plane as well, and deviates from the growth direction (the $z$-axis)
by an angle of $\theta$.

\begin{figure}[t]

\begin{raggedleft}
\includegraphics[width=1\columnwidth]{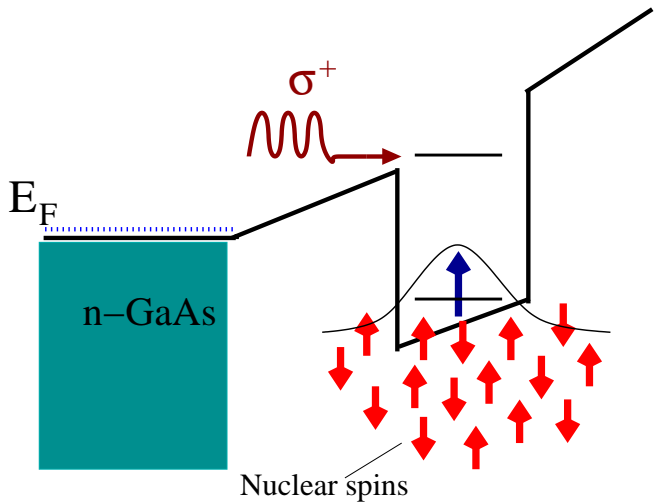}
\par\end{raggedleft}

\caption{\label{fig:ENCElectron-Nuclear-spin-coupled}Schematic of the setup for a typical optical orientation
experiment.\cite{Gammon2001,Bracker2005} The self-assembled quantum dot is embedded in a Schottky diode heterostructure, where the gate voltage
is tuned to allow only one electron orbital state below the Fermi level.  Nuclear spins are polarized by an optically pumped electron captured
in the quantum dot.}

\end{figure}

We base our calculation on the experimental conditions in Refs.~\onlinecite{Braun2006}, \onlinecite{Eble2006},
 \onlinecite{Gammon2001}, \onlinecite{Bracker2005} (see Fig. \ref{fig:ENCElectron-Nuclear-spin-coupled}),
where the SAQD is embedded in a Schottky diode heterostructure, so
that the charged states of the QD can be controlled. In addition,
the gate voltage can be tuned to allow zero or one charge (electron
or hole) on the dot.

In such a system dynamic nuclear spin polarization is realized via
optical pumping. A circularly polarized photon creates an electron-hole
pair, which is then captured in the QD as a negative trion,\cite{Bracker2005,Lai2006}
neutral exciton,\cite{Gammon2001} or positive trion.\cite{Braun2006,Eble2006,Tartakovskii2007}
When the QD contains one electron, this electron is likely spin polarized and can polarize
a nuclear spin through the hyperfine interaction. The probability to realize this spin transfer process depends on
the experimental conditions, such as the type of excitation used (pulsed
or CW) and the initial charged states in the quantum dot ($X^{0}$,
$X^{-}$or $X^{+}$). Therefore, an experimentally determined factor
$f_{e}$ is used to modify the electron-nuclear spin transfer probability
(see \ref{sec:Model} C).

In a finite magnetic field the hyperfine-mediated transfer of polarization
from the electron spin to the nuclear spins has to be assisted by
another process because of the large mismatch of the electron and nuclear
Zeeman energies. In our case we consider phonon-assisted and tunneling-assisted
processes. More specifically, in the Schottky-diode configuration
the confined electron can flip its spin via cotunneling to the external
reservoir (For details see \ref{sub:Cotunneling-processes}). This
cotunneling-assisted spin flip process is efficient at low magnetic
fields. In the high field regions (or in an isolated QD), the electron-phonon
interaction provides the more efficient channel to compensate for
the energy mismatch between electron and nuclear spins (For details
see \ref{sub:Electron-phonon-interaction}).

\section{\label{sec:Model}Model Hamiltonian}

The total Hamiltonian for the nuclear spin polarization scheme we
consider is given as follows, \[
H_{T}=H_{n}+H_{e}+H_{hf},\]
 where $H_{n}$ is the Hamiltonian for the nuclear spins in the quantum
dot, $H_{e}$ is for the electron spin, and $H_{hf}$ is the hyperfine
interaction between the electron and nuclear spins. Below we describe
each of the terms in $H_{T}$ in detail, and discuss the role they
play in the DNSP process.

\subsection{The Hamiltonian of nuclear spin in quantum dots}

In the presence of an external magnetic field (along the $z$-axis),
the Hamiltonian of nuclear spins in our simplified model of a quantum
dot (see Fig.~\ref{fig:sadthe-geometry-of}) is defined as follows,
\begin{eqnarray}
H_{n} & = & H_{n}^{z}+H_{Q}+H_{d},\label{eq:hamiltonian}\end{eqnarray}
 where \begin{eqnarray*}
H_{n}^{z} & = & \sum_{i=1}^{N}\hbar\gamma B_{z}I_{z}^{i},\\
H_{Q} & = & \sum_{i=1}^{N}\frac{eQV_{ZZ}^{i}}{4I(2I-1)}\left[3\left(I_{Z}^{i}\right)^{2}-I(I+1)\right],\\
H_{d} & = & \sum_{i<j}^{N} \frac{\mu_{0}\hbar^{2}\gamma^{2}}{4\pi}
\left[\frac{\mathbf{I}^{i}\cdot\mathbf{I}^{j}}{R_{ij}^{3}}-\frac{3\left(\mathbf{I}^{i}\cdot\mathbf{R}_{ij}\right)
\left(\mathbf{I}^{j}\cdot\mathbf{R}_{ij}\right)}{R_{ij}^{5}}\right].
\end{eqnarray*}
$\quad$$H_{z}^{n}$ represents the nuclear Zeeman energy ($E_{n}^{z}$) where $I_{z}$ is the projection of a nuclear spin along the external
magnetic field, and $\gamma$ is the nuclear gyromagnetic ratio (see Table \ref{tab:elements}).

\begin{table}[t]

\caption{\label{tab:elements} Material parameters used in our calculation.  Nuclear electric quadrupole moment $Q$ and constant $S_{11}$ (which
relates electric field gradient to strain) are taken from Ref.~\onlinecite{Sundfors1974} and Ref.~\onlinecite{Sundfors1976}. Nuclear spin
gyromagnetic ratios are taken from Ref.~\onlinecite{Weast2007}.}

\begin{tabular}{lccc}
 &  &  & \tabularnewline
\hline
\hline
Elements  & In  & Ga  & As\tabularnewline
\hline
nuclear spin $I$  & 9/2  & 3/2  & 3/2\tabularnewline
\noindent electric quadrupole moment Q ($10^{-24}$ cm$^{2}$)  & 0.86  & 0.27  & 0.2\tabularnewline
\noindent $S_{11}$ ($10^{15}$ statcoulombs/cm$^{3}$)  & 16.7  & 9.1  & 13\tabularnewline
\noindent gyromagnetic ratio $\gamma$ ($\mu$eV/T)  & 0.039  & 0.042  & 0.03\tabularnewline
hyperfine constant A ($\mu$eV)  & 56  & 42  & 46\tabularnewline
\hline
\hline
 &  &  & \tabularnewline
\end{tabular}
\end{table}

$H_{Q}$ represents the electric quadrupole interaction ($E_{Q}$),\cite{Abragam1961,Slichter1992} through which the nuclear spins in an
In$_{x}$Ga$_{1-x}$As SAQD couple to the electric field gradients in the crystal lattice. The asymmetric part of the quadrupole interaction is
neglected here because we assume a pancake-shaped QD, where $V_{ZZ}\gg V_{XX},V_{YY}$. Q is the electric quadrupole moment of a nucleus, and $e$
is the elementary charge. $V_{ZZ}$ is the electric field gradient along the principal axis \textit{Z}. $V_{ZZ}=S_{11}e_{ZZ}$,\cite{Dzhioev2007}
where the constant $S_{11}$ is experimentally determined (see in Table \ref{tab:elements}), and $e_{ZZ}$ is the $Z$ component of the strain
tensor, which is approximately 4\% to 8\% for In$_{x}$Ga$_{1-x}$As QDs with different
compositions.\cite{Williamson1999,Grundmann1995,Korkusinski2001,Stier1999,Sheng2005,Yang2008}
The electric field gradients introduced by charged states in the QD\cite{Paget2008}  are at least two orders of magnitude smaller than
 the electric field gradient caused by the broken symmetry of the crystal lattice, therefore we exclude the effect of the former.

$H_{d}$ represents the dipolar interaction between different nuclear spins, where $R_{jk}$ is the distance between the $i$th and the $j$th the
nucleus, $\mu_{0}$ is the free space permeability. The local field $B_{loc}$ that this dipolar coupling produces is around 1 Gauss, and the
dynamical effect of the dipolar interaction is nuclear spin diffusion. In our calculation, we do not deal with the case when the external
magnetic field is smaller than the local field (the smallest external field we use is 10 mT). Furthermore, nuclear spin diffusion is strongly
suppressed in a small QD,\cite{Deng2005a} and its time scale (tens of seconds to minutes\cite{Yusuf2010}) is much longer than the time scale of
DNSP in our study. Therefore we exclude $H_{d}$ in our calculation. Since $H_{d}$ is the only direct interaction between nuclear spins, its
removal significantly simplifies our study: the nuclear spins can now be treated independently from each other. The hyperfine-mediated nuclear
spin interaction\cite{Deng2006,Yao2006,Cywinski2009} is also neglected in our calculation, as the hyperfine interaction is only turned on randomly
for a small fraction of time in experiments, as we will discuss in Section III.C.

\begin{figure}[t]

\begin{centering}
\subfigure[]{\includegraphics[width=0.35\columnwidth]{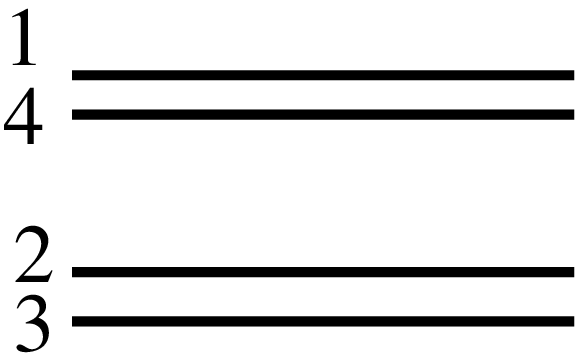}}
$\quad$\subfigure[]{\includegraphics[width=0.35\columnwidth]{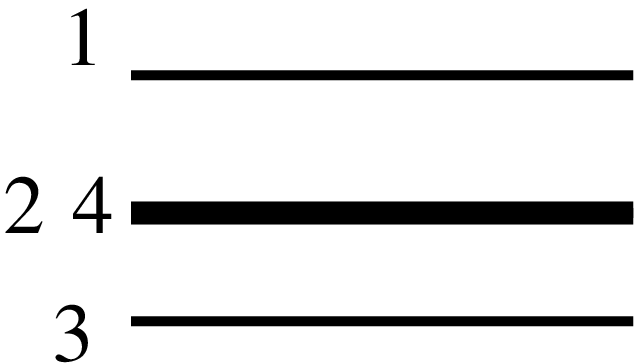}}
\par\end{centering}

\begin{centering}
\subfigure[]{\includegraphics[width=0.35\columnwidth]{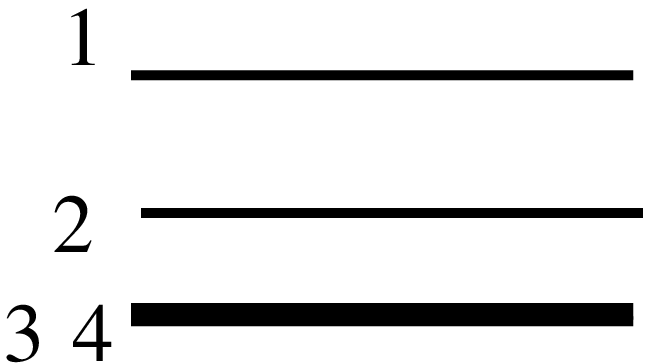}}
$\quad$\subfigure[]{\includegraphics[width=0.35\columnwidth]{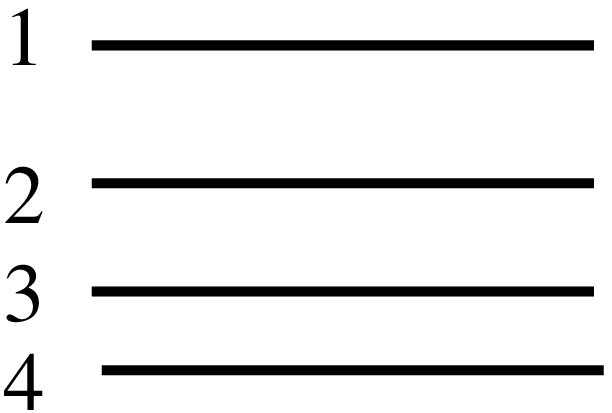}}
\par\end{centering}

\caption{\label{fig:enrgy-levelEnergy-levels}Sketches of the energy-level diagram of a single nuclear spin (not to scale). Due to the presence
of the quadrupole interaction, the eigenstates are generally no longer the eigenvectors of $I_{z}$.  We label the real eigenstates as
\{$\left|1\right\rangle $, $\left|2\right\rangle $, $\left|3\right\rangle $, $\left|4\right\rangle $\}, which are combinations of $I_{z}$
eigenstates with $m_{z}=\{\frac{3}{2},$ $\frac{1}{2},$-$\frac{1}{2},$ -$\frac{3}{2}\}$. (a) Low field situations, when the quadrupole
interaction ($E_{Q}$) is stronger than the nuclear Zeeman effect ($E_{z}^{n}$). (b) 2-4 degeneracy, when the quadrupole interaction and nuclear
Zeeman energy are resonant ($E_{z}^{n}=E_{Q}$). (c) 3-4 degeneracy, when $E_{z}^{n}=2E_{Q}$. (d) High magnetic field region, when the Zeeman
energy is dominant, and the quadrupole interaction is negligible.}

\end{figure}

The nuclear spin Hamiltonian is thus simplified as
 \begin{eqnarray*}
H_{n} & = & \hbar\gamma B_{z}I_{z}\\
 &  & +\frac{eQV_{ZZ}^{}}{4I(2I-1)}\left[I_{Z}^{2}-\frac{1}{3}I\left(I+1\right)\right].\end{eqnarray*}
 In our simplified model of the quantum dot, the principal axis $Z$
for the largest electric field gradient ${V}_{ZZ}$ deviates from
the $z$-axis by an angle $\theta$, and $I_{Z}$ is the projection
of a nuclear spin along the principal axis $Z$. For a pancake-shaped
QD, the angle $\theta$ is generally quite small. Therefore, while
for all the calculations presented in this paper we treat the nuclear
spin Hamiltonian (\ref{eq:hamiltonian}) exactly, for the qualitative
discussion in Section \ref{sec:Calculation} we take a small-angle
approximation and simplify the Hamiltonian (\ref{eq:hamiltonian}).
For example, in the case of $I=3/2$ we obtain, \begin{widetext}

\vspace{5mm}

\noindent \begin{eqnarray}
 &  & \begin{array}{cccc}
\hphantom{1mm}I_{z}=\frac{3}{2} & \hphantom{1m}I_{z}=\frac{1}{2} & \hphantom{1m}I_{z}=-\frac{1}{2}
& \hphantom{1m}I_{z}=-\frac{3}{2}\end{array}\nonumber \\
H_{n}^{z}+H_{QI} & = & \left(\begin{array}{cccc}
\frac{3}{2}E_{z}^{n}+E'_{Q} & \sqrt{3}E_{Q}\theta & \frac{\sqrt{3}}{2}E_{Q}\theta^{2} & 0\\
\sqrt{3}E_{Q}\theta & \frac{1}{2}E_{z}^{n}-E'_{Q} & 0 & \frac{\sqrt{3}}{2}E_{Q}\theta^{2}\\
\frac{\sqrt{3}}{2}E_{Q}\theta^{2} & 0 & -\frac{1}{2}E_{z}^{n}-E'_{Q} & -\sqrt{3}E_{Q}\theta\\
0 & \frac{\sqrt{3}}{2}E_{Q}\theta^{2} & -\sqrt{3}E_{Q}\theta & -\frac{3}{2}E_{z}^{n}+E'_{Q}\end{array}\right),
\label{eq:approxHam}
\end{eqnarray}

\vspace{5mm}

\noindent \end{widetext} where $E'_{Q}=E_{Q}(3\cos^{2}\theta-1)/2\approx E_{Q}$.
When the off-diagonal terms in Eq. (\ref{eq:approxHam}) are small
compared to the diagonal terms, we can construct the new eigenstates
perturbatively. In a non-degenerate case, the nuclear spin eigenstates
are as follows:
\begin{eqnarray}
\Bigl|1\Bigr\rangle & = & \left|+\frac{3}{2}\right\rangle +a\left|+\frac{1}{2}\right\rangle +b\left|-\frac{1}{2}\right\rangle ,\nonumber \\
\Bigl|2\Bigr\rangle & = & \left|+\frac{1}{2}\right\rangle -a\left|+\frac{3}{2}\right\rangle +c\left|-\frac{3}{2}\right\rangle ,\nonumber \\
\Bigl|3\Bigr\rangle & = & \left|-\frac{1}{2}\right\rangle -b\left|+\frac{3}{2}\right\rangle -d\left|-\frac{3}{2}\right\rangle ,\nonumber \\
\Bigl|4\Bigr\rangle & = & \left|-\frac{3}{2}\right\rangle +c\left|+\frac{1}{2}\right\rangle +d\left|-\frac{1}{2}\right\rangle ,
\label{eq:eigenstates}
\end{eqnarray}
where $a=\sqrt{3}E_{Q}\theta/(E_{z}^{n}+2E_{Q})$, $b=\sqrt{3}E_{Q}\theta^{2}/4(E_{z}^{n}+E_{Q})$, $c=\sqrt{3}E_{Q}\theta^{2}/4(E_{z}^{n}-E_{Q})$
and $d=\sqrt{3}E_{Q}\theta/(E_{z}^{n}-2E_{Q})$. For a degenerate case as shown in Fig.~\ref{fig:enrgy-levelEnergy-levels} (b) and (c), the eigenstates have to be solved by directly
diagonalizing the Hamiltonian in Eq.~(\ref{eq:approxHam}). Taking the state 2-4 degeneracy, for example, one would expect a complete mixing
between state 2 and state 4.

\subsection{The Hamiltonian of electron spin in quantum dots}

The effective Hamiltonian of the electron in the quantum dot consists of three main parts: Zeeman splitting, tunnel coupling to the external
reservoir, and the electron-phonon interaction: \[ H_{e}=H_{e}^{z}+H_{T}+H_{ep},\] where
\begin{eqnarray}
H_{e}^{z} & = & -g^{*}\mu_{B}B_{z}S_{z},\nonumber \\
H_{T} & = & \sum_{k\sigma}\varepsilon_{k}n_{k\sigma}+\varepsilon_{0}\sum_{\sigma}n_{\sigma}+Un_{\uparrow}n_{\downarrow}\nonumber \\
 &  & +\sum_{k\sigma}V_{k}\left(c_{k\sigma}^{\dagger}c_{\sigma}+c_{\sigma}^{\dagger}c_{k\sigma}\right),\label{eq:And}\\
H_{ep} & = & \sum_{q\nu}M_{q\nu}(a_{-q\nu}^{\dagger}+a_{q\nu})exp(i\mathbf{q}\cdot\mathbf{r}).\nonumber
\end{eqnarray}
Here $H_{e}^{z}$ represents the electron Zeeman energy, where $g^{*}$ is the electron effective g-factor in In$_{x}$Ga$_{1-x}$As QDs, $\mu_{B}$
is the Bohr magneton, and $S_{z}$ is the $z$ component of the electron spin operator. $H_{T}$ is the Anderson
Hamiltonian,\cite{Smith2005,Schrieffer1966,Mahan2000} suitable for describing the experimental setup in our consideration (see Fig.
\ref{fig:ENCElectron-Nuclear-spin-coupled}), where a QD is tunnel-coupled to the outside Fermi sea. In Eq. (\ref{eq:And}), $c^{\dagger}$ and $c$
represent electron creation and annihilation operators, and $n$ the number operators. We describe a reservoir state with index $k$, energy
$\varepsilon_{k}$, and electron spin index $\sigma$. The single electron energy level in the quantum dot is $\varepsilon_{0}$. U is the on-site
Coulomb interaction and $V_{k}$ is the tunneling matrix element. $H_{ep}$ represents the electron-phonon coupling, where $a_{-q\nu}^{\dagger}$
and $a_{q\nu}$ represent phonon creation and annihilation operators, with quasi-momentum q and branch index $\nu$. We consider both the
deformation potential and piezoelectric potential in $M_{q\nu}$.\cite{Mahan2000}

\subsection{The hyperfine interaction}

In our scheme, nuclear spin polarization (NSP) is pumped by optically oriented electrons via the contact hyperfine interaction:
\begin{eqnarray}
H_{hf}(t) & = & h(t)\sum_{k}^{N}A\left|\Phi(R_{k})\right|^{2}\nonumber \\
 &  & \times\left[I_{z}^{k}S_{z}+\frac{1}{2}\left(I_{+}^{k}S_{-}+I_{-}^{k}S_{+}\right)\right],
 \label{eq:hyperfine}
\end{eqnarray}
where $A$ is the hyperfine coupling constant (see Table \ref{tab:elements}). $N$ is approximately $10^{4}$ in an In$_{x}$Ga$_{1-x}$As SAQD.
$\Phi(R_{k})$ is the electron wave function at the $k$th nucleus site, which is Gaussian for harmonic confinement. For our calculations
presented in this paper, we take $\left|\Phi(R_{k})\right|^{2}$ as $1/N$, effectively assuming a constant electron wave function in the QD. This
assumption makes the definition of nuclear spin polarization well-defined, while in the case of a Gaussian wave function the calculation of
overall NSP depends on where the dot is truncated, as the edge of the dot would generally be only slightly polarized. $h(t)$ is a random
function depending on the experimental procedures and conditions. While we are not going to describe the full details of each pumping scheme, the
nature of $h(t)$ depends on whether the system is in the trion or neutral exciton regime. For example, for the $X^{+}$ scheme, $h(t)$ is
dependent on the trapping and recombination of the electron;\cite{Braun2006} while for $X^{-}$ scheme it is dependent on the exciton
recombination and electron tunneling time;\cite{Bracker2005,Maletinsky2007} and for $X^{0}$ it is dependent on the exciton recombination
time.\cite{Gammon2001} The fraction $f_{e}$ is defined as the mean value of this temporal function $h(t)$, and it represents the fraction of the
time when only one electron is left in the quantum dot and the hyperfine interaction is {}``turned on'', so that the electron-nuclear spin flip-flop
 can be realized. We use 0.035 for $f_{e}$ in our calculations, based on experimental observations.\cite{Maletinsky2007} The small value of
$f_{e}$, together with the fact that $h(t)$ is random in time to a degree, justify our approximation of neglecting higher-order effects of the
hyperfine interaction throughout our calculations.

\begin{figure}[t]
\includegraphics[width=1\columnwidth]{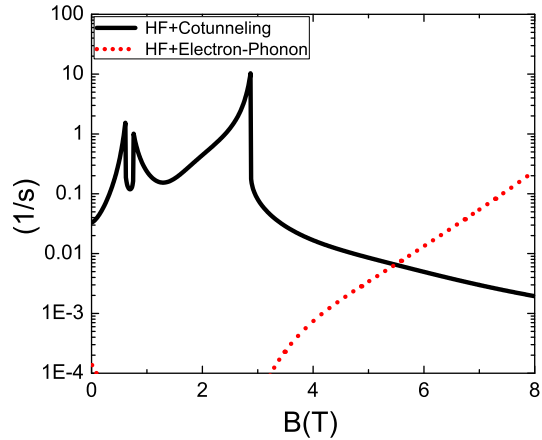}
\caption{\label{fig:pumprate}(color online). Comparison of cotunneling-assisted nuclear spin pumping rates with phonon-assisted rates. Here the Overhauser field
is anti-parallel to the external magnetic field, and the angle between the largest electric field gradient and the external magnetic field is
$2^{\circ}$ (we use these conditions in the following figures, unless otherwise noted). The cotunneling-assisted spin flip process is more efficient at low to intermediate magnetic fields (B < 5 T). For higher
fields, the phonon-assisted spin flip process is more efficient.}

\end{figure}

\begin{figure}[t]

\begin{centering}
\includegraphics[width=1\columnwidth]{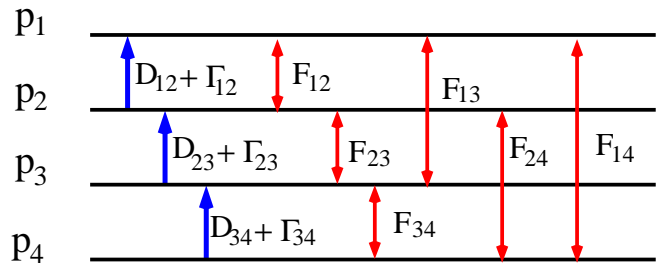}
\par\end{centering}

\caption{\label{fig:Hyperfne-induced-transition-among}(color online). Transitions induced by the hyperfine interaction among nuclear spin states
with $I=3/2$.  The blue one-way arrows represent the transitions induced by the flip-flop terms in Eq.~(\ref{eq:hyperfine}), and are responsible
for pumping of the nuclear spins by the trapped electron spin.  For an open-dot system as shown in
Fig.~\ref{fig:ENCElectron-Nuclear-spin-coupled}, the cotunneling-assisted spin flip rate $D$ is dominant at low to intermediate magnetic fields.
At high fields or in an isolated dot, the phonon-assisted pumping rate $\Gamma^{ph}$ is dominant.  The red two-way arrows
represent transitions due to the mixture of different nuclear spin states, which originate from the $I_z S_z$ term in Eq.~(\ref{eq:hyperfine}).
These processes are responsible for depolarization.}

\end{figure}

Depending on the helicity of the optical excitation ($\sigma^{+}/\sigma^{-}$) relative to the applied external magnetic field, and through the
flip-flop terms in the hyperfine interaction, the electron can pump nuclear spins either to the highest-energy spin state or the lowest-energy
state. These spin flip-flops are responsible for pumping the nuclear spins in the NSP process (blue one-way arrows in
Fig.~\ref{fig:Hyperfne-induced-transition-among}). In subsection \ref{sub:Cotunneling-processes} and \ref{sub:Electron-phonon-interaction} we
calculate the pumping rates based on the specific physical processes involved.

\subsubsection{\label{sub:Cotunneling-processes}Cotunneling-assisted spin flip
processes}

The electron spin in the QD can interact with an electron spin in
the Fermi sea via cotunneling processes, so that the electron Zeeman levels
are broadened. The spin flip probability and level broadening can
be calculated applying Schrieffer-Wolf transformation to Eq. (\ref{eq:And}).\cite{Schrieffer1966,Warburton2000,Smith2005,Dreiser2008}
For each Zeeman level, \[
\rho(\varepsilon_{i})=\frac{1}{2\pi}\frac{\Gamma_{c}}{(\varepsilon_{i}-\varepsilon_{0})^{2}+\Gamma_{c}^{2}/4},\]
 where $\Gamma_{c}$ is the level broadening due to the cotunneling
processes.

The probability of electron nuclear spin flip-flop processes increases
when the overlap between the two broadened electron Zeeman states
increases. Therefore the cotunneling-assisted spin flip-flop is more
important at low magnetic fields. The transition rate between an initial
state $\left|\uparrow i\right\rangle $ and a final state $\left|\downarrow j\right\rangle $
can be calculated with the Fermi Golden Rule as,
\begin{eqnarray}
D_{i,j} & = & \frac{2\pi}{\hbar}f_{e}\left(\frac{A}{N}\right)^{2}\left|\left\langle \downarrow j
\right|I_{+}S_{-}\left|\uparrow i\right\rangle \right|^{2}\nonumber \\
 &  & \times\int\rho(\varepsilon_{i})\rho(\varepsilon_{f})\delta(\varepsilon_{f}-\varepsilon_{i}
 -\vartriangle E)d\varepsilon_{i}d\varepsilon_{f}\nonumber \\
 & \approx & \left(\frac{A}{N}\right)^{2}\frac{2f_{e}\tau_{c}\left|\left\langle \downarrow j
 \right|I_{+}S_{-}\left|\uparrow i\right\rangle \right|^{2}}{\hbar^{2}+\tau_{c}^{2}\left(-g^{*}
 \mu_{B}B_{z}+\delta_{n}\right)^{2}},\label{eq:pump1}
\end{eqnarray}
 where $\delta_{n}=2A\left\langle I_{z}\right\rangle $ stands for
the Overhauser shift, and the $+$ sign in front of the Overhauser
field is due to our choice here that the external magnetic field is
anti-parallel to the Overhauser field. For parallel fields, $-\delta_{n}$
should be used. The correlation time, $\tau_{c}=1/\Gamma_{c}$, describes
the broadening of QD electron states due to cotunnueling processes.
It is estimated for a typical Schottky structure to be $\sim20$ ns.\cite{Smith2005}
At low to intermediate magnetic fields, the cotunneling-assisted spin
flip processes are the most efficient in building up NSP in the QD,
as shown in Fig. \ref{fig:pumprate}.


\subsubsection{\label{sub:Electron-phonon-interaction}Phonon-assisted spin flip
processes}

For an isolated dot, or a dot described in \ref{sub:Cotunneling-processes} in higher field regions, the cotunneling-assisted spin flip processes
become less efficient due to the larger electronic Zeeman splitting. Now the phonon-assisted spin flip processes give the most efficient DNSP
channel.
The pumping rates due to the phonon-assisted spin flip processes are,
\begin{eqnarray*}
\Gamma_{i,j}^{ph} & = & \frac{2\pi}{\hbar}\sum_{q\nu}\left|T_{ep}\right|^{2}\\
 &  & \times\left[\bar{n}_{q\nu}\delta(\hbar sq-E_{z}^{e})+(\bar{n}_{q\nu}+1)\delta(E_{z}^{e}+\hbar sq)\right]\\
 & =N_{q} & \left[f_{def}(E_{z}^{e})+f_{piezo}(E_{z}^{e})\right].\end{eqnarray*}
 where \begin{eqnarray*}
T_{ep} & = & \sum_{l\neq m}\frac{\left\langle m\downarrow j\right|H_{hf}\left|l\uparrow i\right\rangle \left\langle l
\right|H_{ep}\left|m\right\rangle }{E_{m}-E_{l}+E_{z}^{e}}\\
 &  & +\frac{\left\langle m\right|H_{ep}\left|l\right\rangle \left\langle l\downarrow j\right|H_{hf}\left|m\uparrow i
 \right\rangle }{E_{m}-E_{l}-E_{z}^{e}},\end{eqnarray*}
 \begin{eqnarray*}
&&f_{def}(E_{z}^{e})\approx\frac{l_{0}^{2}}{24\pi}\frac{\Xi^{2}f_{e}}{\rho s}\left(\frac{A}{N}\right)^{2}\frac{
\left(E_{z}^{e}\right)^{5}}{\left(\hbar s\right)^{6}}\left|\left\langle \downarrow j\right|I_{+}S_{-}
\left|\uparrow i\right\rangle \right|^{2}\\
 &  & \left\{ \left[\frac{1}{\hbar\Omega_{-}}\left(1+\frac{E_{z}^{e}}{\hbar\Omega_{-}}\right)+\frac{1}{\hbar\Omega_{+}}
 \left(1-\frac{E_{z}^{e}}{\hbar\Omega_{+}}\right)\right]^{2}\right.\\
 &  & \left.+\left[\frac{1}{\hbar\Omega_{+}}\left(1+\frac{E_{z}^{e}}{\hbar\Omega_{+}}\right)+\frac{1}{\hbar\Omega_{-}}
 \left(1-\frac{E_{z}^{e}}{\hbar\Omega_{-}}\right)\right]^{2}\right\} ,\\
&&f_{piezo}(E_{z}^{e})\approx\frac{l_{0}^{2}}{30\pi}\frac{(ee_{14})^{^{2}}f_{e}}{\rho s}\left(\frac{A}{N}\right)^{2}
\frac{\left(E_{z}^{e}\right)^{3}}{\left(\hbar s\right)^{4}}\left|\left\langle \downarrow j\right|I_{+}S_{-}
\left|\uparrow i\right\rangle \right|^{2}\\
 &  & \left\{ \left[\frac{1}{\hbar\Omega_{-}}\left(1+\frac{E_{z}^{e}}{\hbar\Omega_{-}}\right)+\frac{1}{\hbar\Omega_{+}}
 \left(1-\frac{E_{z}^{e}}{\hbar\Omega_{+}}\right)\right]^{2}\right.\\
 &  & \left.+\left[\frac{1}{\hbar\Omega_{+}}\left(1+\frac{E_{z}^{e}}{\hbar\Omega_{+}}\right)+\frac{1}{\hbar\Omega_{-}}
 \left(1-\frac{E_{z}^{e}}{\hbar\Omega_{-}}\right)\right]^{2}\right\} .
 \end{eqnarray*}
The initial state is $\left|m\uparrow i\right\rangle $ and the final state is $\left|m\downarrow j\right\rangle $, where $i$ and $j$ represent
nuclear spin eigenstates. $m$ stands for the initial orbital state (QD s orbital). $\bar{n}_{q\nu}$ is the Bose-Einstein distribution for
phonons with momentum q and phonon branch $\nu$ at temperature $T$. We consider both phonon absorption and emission processes, depending on the
direction of the total magnetic field. $N_{q}=\bar{n}_{q\nu}$ for phonon emission processes, while $N_{q}=\bar{n}_{q\nu}+1$ for phonon
absorption processes. $f_{def}(E_{z}^{e})$ is obtained from the deformation potential term and $f_{piezo}(E_{z}^{e})$ is obtained from the
piezoelectric interaction. $T_{ep}$ is the transition amplitude for the phonon-assisted spin flip processes. Here the hyperfine interaction not
only induces electron nuclear spin flip-flop, but also mixes electron spin and orbital degrees of freedom.\cite{Erlingsson2002} The electron
spin-up (-down) state in the $m$ orbital is mixed with the electron spin-down (-up) state from a higher orbital state $l$. $E_{m}$ and $E_{l}$
represent the energies of the $m$ and $l$ Fock-Darwin orbital states. For simplicity, we only consider the phonon emission/absorption between s
orbital and p orbital states (we do not anticipate the inclusion of contributions from higher orbital states to qualitatively alter our
results). $\Xi$ is the deformation potential constant. The piezoelectric constant is denoted as $ee_{14}=2\times10^{-10}$ J/m, sound speed:
$s=3\times10^{3}$ m/s, electron density in InAs: $\rho=5.7\times10^{3}$ Kg/m$^{3}$. In the presence of an external magnetic field, Fock-Darwin
energy levels can be represented as $\hbar\Omega_{\pm}=\hbar\Omega\pm\hbar\omega_{c}/2$, where $\hbar\Omega$ is the electronic confinement in
the QD and is about 30 meV in the type of QD we consider. The cyclotron frequency is $\omega_{c}=eB/m^{*}$ where $m^{*}=0.023\: m_{0}$ is the
effective electron mass in the InAs QDs. $l_{0}$ is the lateral dimension of the QD. The transition rate due to the deformation potential is
proportional to the fifth power of the electronic Zeeman splitting, while the contribution from the piezoelectric interaction is proportional to
the third power of the electronic Zeeman spitting.

\subsubsection{\label{sub:Depolarization-Channel}Strain-induced depolarization}

Due to the strain-induced quadrupole interaction, where the principal
axis $Z$ is generally not parallel to the external field direction
$z$, the nuclear spin eigenstates are a mixture of $I_{z}$ eigenstates.
This means that the non flip-flop term in Eq. (\ref{eq:hyperfine}),
$I_{z}^{k}S_{z}$, can now induce transitions between different nuclear
spin states and cause NSP (see red two-way arrows in Fig.~\ref{fig:Hyperfne-induced-transition-among}).
Since the energy transfer between these nuclear spin states is generally
much less than cotunneling energy (0.033$\mu$eV\cite{Smith2005}),
this process is not limited by energy conservation considerations.
These depolarization rates can be calculated by the Fermi Golden Rule,
\begin{equation}
F_{ij}=\left(\frac{A_{As}}{N}\right)^{2}\frac{2\tau_{c}f_{e}\left|\left\langle i\right|I_{z}\left|j\right\rangle
\right|^{2}}{\hbar^{2}+\tau_{c}^{2}\left(E_{i}^{n}-E_{j}^{n}\right)^{2}},\label{eq:depo}
\end{equation}
where $i$ and $j$ represent nuclear spin eigenstates, and $E_{i}^{n}$ and $E_{j}^{n}$ stand for the nuclear spin eigen-energies.

\subsection{Master equation of population \label{sub:Master-equation-of}}

Depending on the helicity of the excitation photon ($\sigma^{+}/\sigma^{-}$),
the electron can pump nuclear spins either to the higher-energy spin
states or the lower-energy states. Take for example nuclear spins
being pumped to the highest-energy nuclear spin state, as shown in
Fig.~\ref{fig:Hyperfne-induced-transition-among}, the average NSP
can be evaluated by the master equation of population [see Eq.~(\ref{eq:mastereq})],\cite{Abragam1961}
which is determined by the balance between the pumping and depolarization
channels.
\begin{equation}
\frac{dp_{i}}{dt}=\sum_{j\neq i}W_{j,i}p_{j}-\sum_{j\neq i}W_{i,j}p_{i},\label{eq:mastereq}
\end{equation}
 where $W_{j,i}\neq W_{i,j}$ represents the total transition rate
between the i and j states, and $p_{i}$ represents the nuclear spin
population at $i$th state with $i=1,2,3,4$.

\[
\frac{d}{dt}\left[\begin{array}{c}
p_{1}\\
p_{2}\\
p_{3}\\
p_{4}\end{array}\right]=M\,\left[\begin{array}{c}
p_{1}\\
p_{2}\\
p_{3}\\
p_{4}\end{array}\right],\]
 where \begin{widetext}

\vspace{6mm}

$M=\left[\begin{array}{cccc}
-(F_{12}+F_{13}+F_{14}) & (D'_{12}+F_{12}) & F_{13} & F_{14}\\
F_{12} & -(D'_{12}+F_{12}+F_{23}+F_{24}) & (F_{23}+D'_{23}) & F_{24}\\
F_{13} & F_{23} & -(D'_{23}+F_{23}+F_{34}+F_{13}) & (D'_{34}+F_{34})\\
F_{14} & F_{24} & F_{34} & -(D'_{34}+F_{34}+F_{24}+F_{14})\end{array}\right]\,.$

\vspace{6mm}

\end{widetext} Here $D'_{ij}=D_{ij}+\Gamma_{ij}$ is the total pumping
rate. The system is highly nonlinear due to the population dependence
of the pumping rate. The steady state nuclear spin polarization $\left\langle I_{z}\right\rangle =\sum_{i}p_{i}\left\langle I_{i}\right\rangle $
has to be calculated self-consistently.

\section{\label{sec:Calculation}Results of nuclear spin polarization}

Recent experiments have achieved NSP in In$_{x}$Ga$_{1-x}$As QDs
at various magnetic fields. NSP of $10\sim20\%$ is created in In$_{0.9}$Ga$_{0.1}$As QDs at below 1 T,\cite{Lai2006}
40\% in In$_{0.6}$Ga$_{0.4}$As at approximately 2 T,\cite{Eble2006,Tartakovskii2007}
and 80\% in In$_{0.9}$Ga$_{0.1}$As at 5 T.\cite{Maletinsky2008}
To better understand the differences in these results, we calculate
the NSP in In$_{0.25}$Ga$_{0.75}$As (QD1), In$_{0.6}$Ga$_{0.4}$As
(QD2), and In$_{0.9}$Ga$_{0.1}$As (QD3) quantum dots at various
magnetic fields. The default temperature for our calculations is 4
K, unless otherwise identified.

The $z$ component strain tensor $e_{ZZ}$ is a good indication of
strain strength in our simplified model of the QD. As shown in Fig.~\ref{fig:sadthe-geometry-of},
the largest electric field gradient is ${V}_{ZZ}$, which is proportional to $e_{ZZ}$. Near the surface
of a pancake-like pure InAs QD embedded in GaAs, $e_{ZZ}$ ranges
between $4\sim8\%$.\cite{Grundmann1995,Yang2008,Williamson1999,Korkusinski2001}
Inside, the strain is distributed more or less evenly, and $e_{ZZ}$
is 3.5\%, which is about half of the lattice mismatch between InAs
and GaAs (7\%). To give a quantitative estimate, we assume that the
strain information is completely contained in the $e_{ZZ}$ tensor
element, which reflects the calculated lattice mismatch in dots with
In$_{x}$Ga$_{1-x}$As compositions.\cite{Migliorato2002} By interpolation,
we estimate $e_{ZZ}$ in QD1, QD2 and QD3 to be 2.5\%, 4.3\%, and
6.3\% respectively.

It is worth noting that the electron $g$ factor in a In$_{x}$Ga$_{1-x}$As
self-assembled quantum dot depends on the strain strength. \cite{Pryor2006}
Based on the $g$ factors given in experiments,\cite{Bayer2002,Eble2006,Maletinsky2007}
we take the $g$ factors in QD1, QD2, and QD3 to be -0.6, -0.7, and
-0.8 respectively.

\subsection{Single Arsenic nuclear spin polarization}

To understand nuclear spin polarization in different compositions
of In$_{x}$Ga$_{1-x}$As quantum dots, we first start with the nuclear
spin polarization of As. 
The simplest case is the high magnetic field region, as shown in
Fig.~\ref{fig:enrgy-levelEnergy-levels}(d), where the nuclear spin
Zeeman energy is much larger than the quadrupole splitting. Here nuclear
spin eigenstates are close to the eigenstates of $I_{z}$, so that
$\left\langle i\right|I_{z}\left|j\right\rangle \thickapprox0$. Accordingly,
the depolarization rates [see Eq. (\ref{eq:depo})] between nuclear
spin states are approximately zero. Therefore nuclear spins can be
pumped to the highest nuclear spin state, and nearly full nuclear
spin polarization can be obtained.

\begin{figure}[t]
\begin{centering}
\includegraphics[width=1\columnwidth]{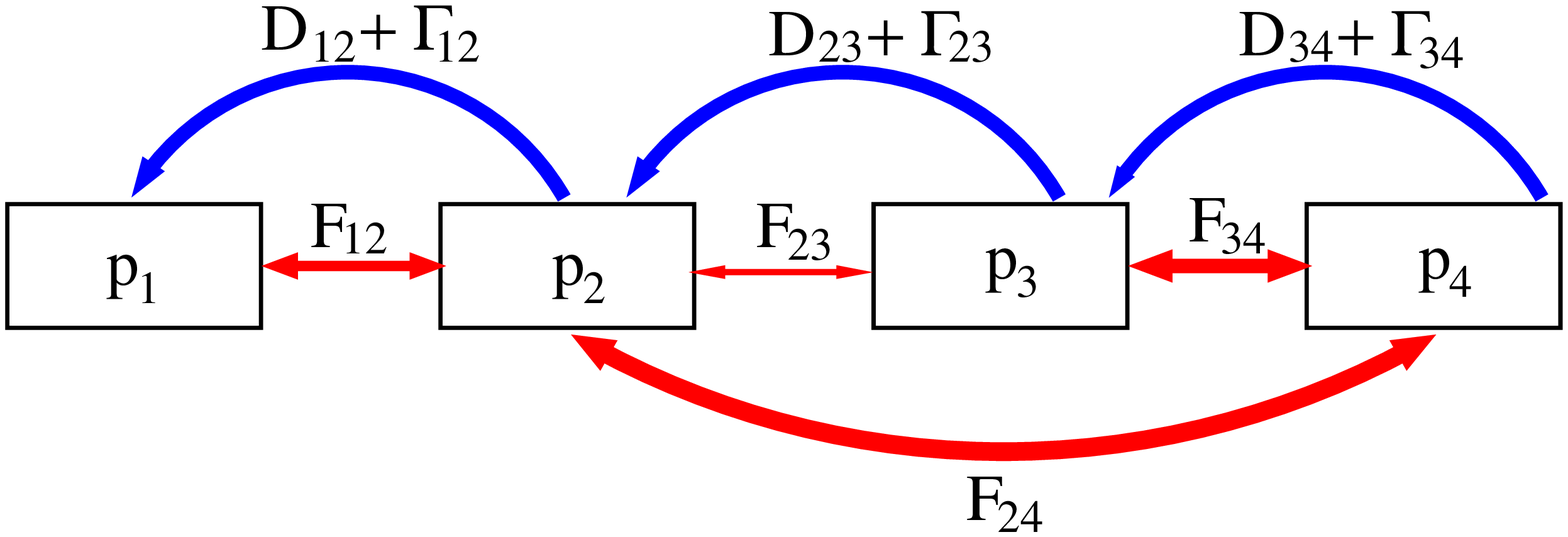}
\par\end{centering}

\caption{\label{fig:depolarization-schemePictorial-explanation}(color online). Sketch of possible transitions between nuclear spin states. 
The blue one-way arrows represent the pumping channels, and the red two-way arrows represent the depolarization channels. 
The thickness of the arrows qualitatively represents the strength of the corresponding transition.}
\end{figure}

In low to intermediate field regions, the calculation for NSP becomes
more complicated since the mixing between different nuclear spin Zeeman
states become stronger than the higher-field case. The physical picture
of various possible transitions is given in Fig.~\ref{fig:depolarization-schemePictorial-explanation}.

\begin{figure}[t]
 \includegraphics[width=1\columnwidth]{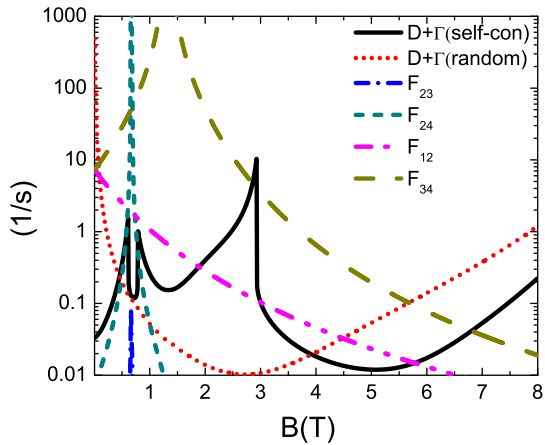}

\caption{\label{fig:all-rates}(color online). An example of pumping and depolarization rates in QD3. $F_{13}$ and $F_{14}$
do not appear here since they are smaller than $10^{-5}/s$. The state 2-4 degeneracy (where the nuclear Zeeman and quadrupole splittings
coincide) occurs at approximately 0.67 T, where $F_{24}$ and $F_{23}$ reach their peaks. The pumping rates $D+\Gamma$(self-con) are calculated
to self-consistency. In other words, these are pumping rates when nuclear polarization is already built up. $D+\Gamma$(random) are the initial
pumping rates with a random distribution of nuclear spins
 before the Overhauser field builds up.}

\end{figure}

According to Fig.~\ref{fig:depolarization-schemePictorial-explanation},
in the absence of all the depolarization channels, nuclear spins can
always be pumped to the highest spin state, and thus full NSP can
be obtained. In the absence of $F_{12}$, no matter how strong other
depolarization channels ($F_{23}$, $F_{24}$, and $F_{34}$) are,
the nuclear spins can still be pumped to state 1 eventually, and thus
become fully polarized. Once $F_{12}$ is turned on, and in combination
with $F_{23}$ or $F_{24}$, the pumped nuclear spins in state 1 can
now leak back to state 3 or 4, and full polarization cannot be achieved.
In other words, $F_{12}$ is the key to depolarization. As an example,
we plot all the depolarization rates and pumping rates in Fig.~\ref{fig:all-rates}.

\begin{figure*}[t]
\begin{centering}
\includegraphics[clip,width=2\columnwidth,height=1.5\columnwidth]{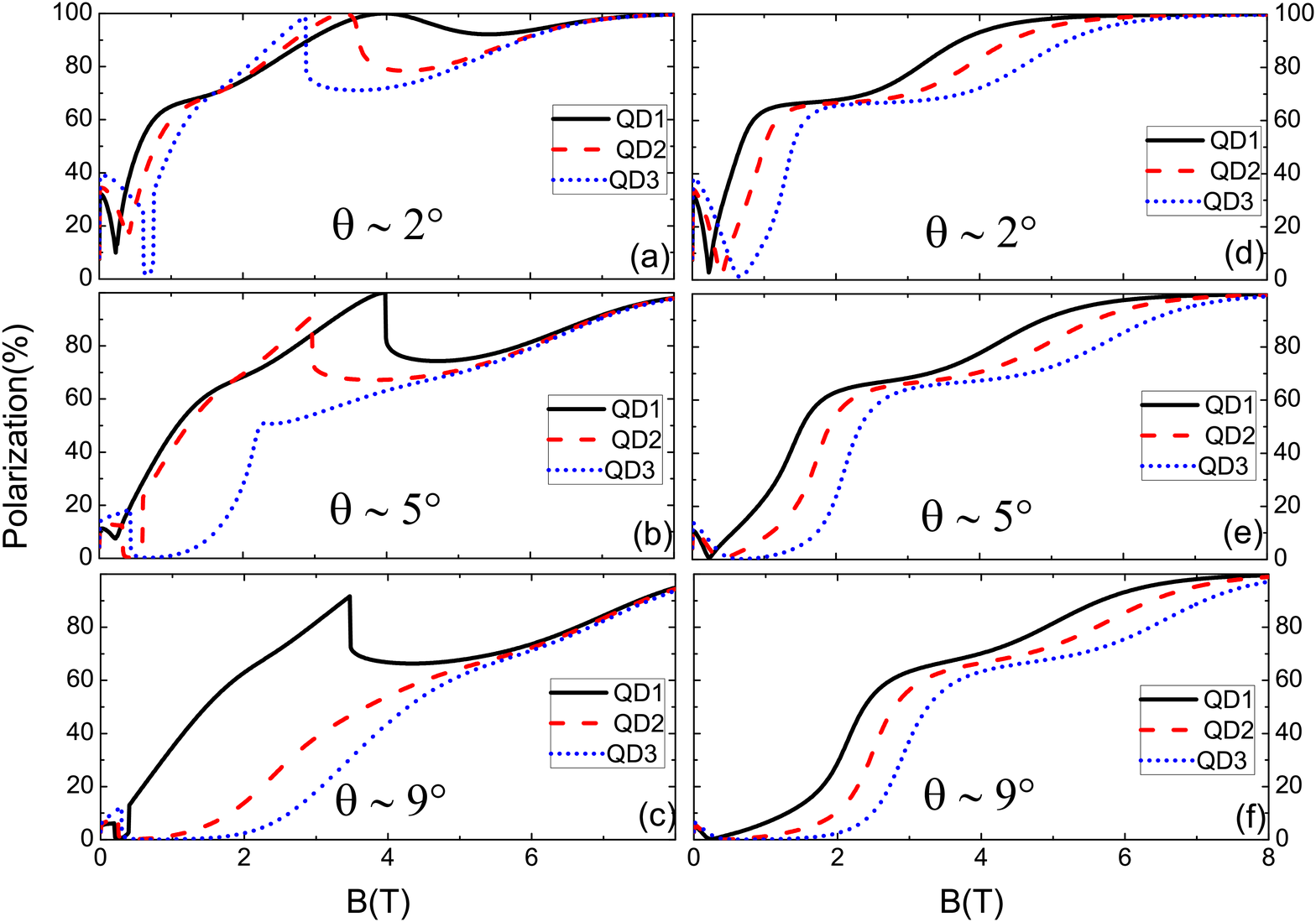}
\par\end{centering}

\caption{\label{fig:DNSP-in-AsNuclear}(color online). Nuclear spin polarization of As in three different In$_{x}$Ga$_{1-x}$As quantum dots, QD1, QD2 and QD3.
The strain strengths in the dots are approximately proportional to the In concentration $x$. The solid, dashed and dotted lines stand for the NSP
in QD1, QD2 and QD3 with $x=0.25$, $0.6$ and $0.9$ respectively. $\theta$ represents the angle between the field gradient and the external
magnetic field (the $z$-axis). Here $\theta = 2^{\circ}$, $5^{\circ}$ and $9^{\circ}$ are considered.  Panels (a), (b) and (c) show As nuclear
spin polarizations at various strain conditions when the Overhauser field is anti-parallel to the external magnetic field. Panels (d), (e) and
(f) show the parallel case.  Overall, stronger strain or greater angle between the field gradient and growth direction suppresses the nuclear
spin polarization.}

\end{figure*}

In the two limits where $E_{z}^{n}>E_{QI}$ and $E_{z}^{n}<E_{QI}$,
the state spectrum is mostly Zeeman-like [Fig.~\ref{fig:enrgy-levelEnergy-levels}(d)]
or quadrupole-like [Fig.~\ref{fig:enrgy-levelEnergy-levels}(a)].
In these cases $F_{23}$ and $F_{24}$ are very small compared to
$F_{12}$ (see Fig.~\ref{fig:all-rates}), so that populations pumped
into states 1 and 2 cannot leak to states 3 and 4. We can now simplify
the four-level problem to a two-level problem, and find the steady
state solution from the following equations:
\begin{eqnarray*}
\dot{p{}_{1}} & = & -F_{12}\: p_{1}+(D_{12}+\Gamma_{12}+F_{12})\: p_{2}=\:0,\\
p_{1} & + & p_{2}\:=\:1\,.
\end{eqnarray*}
 When $E_{z}^{n}<E_{QI}$, the cotunneling-assisted spin flip
transition $D_{12}$ dominates, while when $E_{z}^{n}>E_{QI}$ the phonon-assisted
spin flip transition $\Gamma_{12}$ is dominant, especially above
5 T (see Fig.~\ref{fig:pumprate}). At fields lower than approximately 5 T, the average nuclear spin polarization
can be expressed as follows,
\begin{eqnarray}
\left\langle I_{z}\right\rangle  & \approx & \frac{1}{2}+\frac{1}{1+\frac{F_{12}}{D_{12}+F_{12}}},\label{eq:ana1}
\end{eqnarray}
  while at higher fields
\begin{eqnarray}
\left\langle I_{z}\right\rangle  & \approx & \frac{1}{2}+\frac{1}{1+\frac{F_{12}}{\Gamma_{12}+F_{12}}}.\label{eq:ana2}
\end{eqnarray}
 From Eq.~(\ref{eq:eigenstates}) and Eq.~(\ref{eq:depo}),
when $E_{z}^{n}>E_{QI}$, $F_{12}$ is approximately
\begin{eqnarray*}
F_{12} & = & \left(\frac{A_{As}}{N}\right)^{2}\ \frac{6\tau_{c}f_{e}\theta^{2}(\frac{E_{Q}}{E_{z}^{n}})^{2}}{\hbar^{2}+\tau_{c}^{2}{E_{z}^{n}}^{2}},
\end{eqnarray*}
 and when $E_{z}^{n}<E_{QI}$,
\begin{eqnarray*}
F_{12} & = & \left(\frac{A_{As}}{N}\right)^{2}\frac{6\tau_{c}f_{e}\theta^{2}}{\hbar^{2}+\tau_{c}^{2}{E_{z}^{n}}^{2}}.
\end{eqnarray*}
 Equations (\ref{eq:ana1}) and (\ref{eq:ana2}) can give a very good
qualitative explanation to our calculations. At around 1.5 to 2 T,
as shown in Fig.~\ref{fig:DNSP-in-AsNuclear}(a) and (d), $D_{12}$
is at least one order of magnitude smaller than $F_{12}$, and the
resulting $\left\langle I_{z}\right\rangle $ from Eq.~(\ref{eq:ana1})
is $\sim1$ (NSP is $~67\%$), and is independent of field gradients
of different angles and strengths (which determine $F_{12}$).

\begin{figure}[t]
\begin{centering}
\includegraphics[width=1\columnwidth]{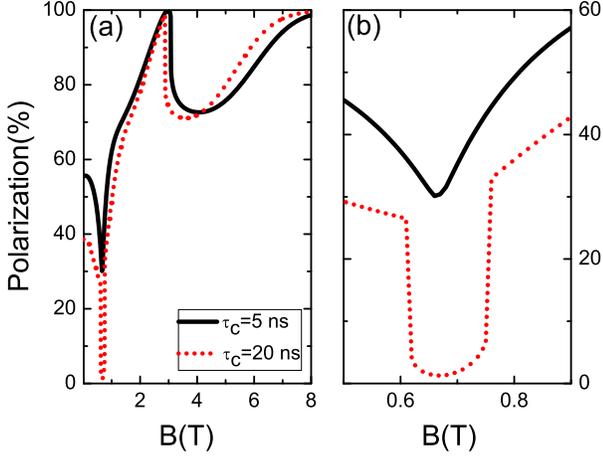}
\par\end{centering}

\caption{\label{fig:cotunneling}(color online). Field dependence of As nuclear spin polarization at different cotunneling rates. As
shown in panel (a), the decrease of the cotunneling time constant ($\tau_{c}$), or the increase of the cotunneling energy, enhances the overlap
of the electronic energy levels, specifically when the external magnetic field is below approximately 5 T. The cross-section of the hyperfine
flip-flop processes is increased, and the resulting nuclear spin polarization increases accordingly.  Panel (b) is a zoom-in of panel (a) near
the 2-4 degeneracy.}

\end{figure}

There are some general trends in the NSP as evident in Eq.~(\ref{eq:ana1})
(we focus on the regime of $B<5$ T for our qualitative discussion
in the following paragraph), where $\left\langle I_{z}\right\rangle $
depends only on the ratio of depolarization to polarization $F_{12}/D_{12}$:
 \begin{eqnarray*}
\left\langle I_{z}\right\rangle  & \approx & \frac{1}{2}+\frac{1}{1+\frac{F_{12}/D_{12}}{1+F_{12}/D_{12}}}\,.
\end{eqnarray*}
 For $E_{z}^{n}>E_{QI}$,
\begin{eqnarray}
\frac{F_{12}}{D_{12}} & = & \frac{\theta^{2}(\frac{E_{Q}}{E_{z}^{n}})^{2}\biggl[(\frac{\hbar}{\tau_{c}})^{2}+\left(-g^{*}\mu_{B}B_{z}+\delta_{n}\right)^{2}\biggr]}{\left[(\frac{\hbar}{\tau_{c}})^{2}+(E_{z}^{n})^{2}\right]}\,;\label{eq:gamma12}
\end{eqnarray}
 and for $E_{z}^{n}<E_{QI}$,
\begin{eqnarray}
\frac{F_{12}}{D_{12}} & \approx & \frac{\theta^{2}\biggl[(\frac{\hbar}{\tau_{c}})^{2}+\left(-g^{*}\mu_{B}B_{z}+\delta_{n}\right)^{2}\biggr]}{(\frac{\hbar}{\tau_{c}})^{2}}.\label{eq:gamma13}
\end{eqnarray}
 Notice that for $E_{z}^{n}>E_{QI}$ the ratio $F_{12}/D_{12}$ is
proportional to the square of $E_{Q}$/$E_{z}^{n}$ and $\theta$.
When $E_{Q}$ or $\theta$ increases, $F_{12}/D_{12}$ increases, and
the average NSP $\left\langle I_{z}\right\rangle $ will decrease.
This is illustrated in the overall trends of Fig.~\ref{fig:DNSP-in-AsNuclear}.
Likewise, in the regime of $E_{z}^{n}<E_{QI}$, the ratio $F_{12}/D_{12}$
is proportional to $\theta^{2}$ and is a function of $\tau_{c}$.
Now when $\theta$ increases, $\left\langle I_{z}\right\rangle $
decreases, again shown in Fig.~\ref{fig:DNSP-in-AsNuclear}. Furthermore,
when the second term of Eq.~(\ref{eq:gamma13}) is greater than the
first (i.e. the electronic Zeeman energy is greater than the cotunneling
energy), $F_{12}/D_{12}\propto\tau_{c}^{2}$, so that $\left\langle I_{z}\right\rangle $
decreases when $\tau_{c}$ increases. As shown in Fig.~\ref{fig:cotunneling},
the decrease of the cotunneling time constant ($\tau_{c}$) enhances
the overlap of the electronic energy levels, and increases the cross-section
of the hyperfine flip-flop processes. Therefore the resulting NSP
increases, as shown in Fig.~\ref{fig:cotunneling}.

\begin{figure}[t]
\begin{centering}
\includegraphics[width=1\columnwidth]{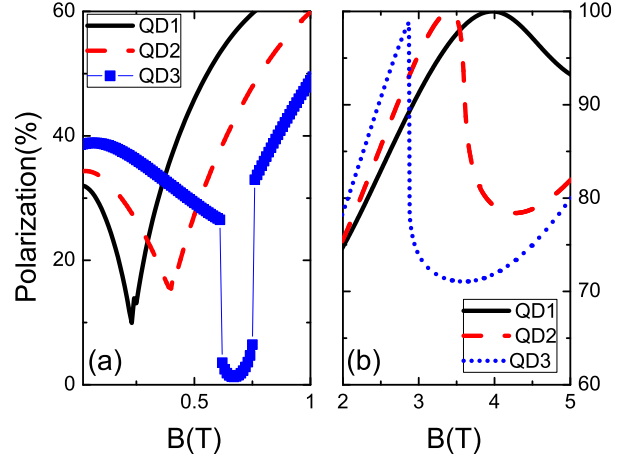}
\par\end{centering}

\caption{\label{fig:zoomin}(color online). The As nuclear spin polarization near 2-4 degeneracy [Panel (a)] and near the peak polarization [Panel (b)].  The two
panels are detailed views of Fig.~\ref{fig:DNSP-in-AsNuclear}(a) in these two regions.  Near 2-4 degeneracy NSP depends on the ratio
$F_{12}/D_{12}$, as shown in Eq.~(\ref{eq:ana2}).  In QD1 and QD2, $F_{12}/D_{12} \sim 1$, and the NSP is above 10\% even at the degeneracy
point. In QD3, $F_{12}/D_{12} \gg 1$, NSP is suppressed at the degeneracy and the field dependence of NSP becomes abrupt.  The NSP peaks in
panel (b) around $3\sim4$ T occur when the Overhauser field cancels out the external mangetic field. These peaks shift to the lower magnetic
field and lower nuclear polarization in stronger strain (see detail in text).}

\end{figure}

The nonlinear nature of our system becomes most prominent when the
Overhauser field is anti-parallel to the external magnetic field, as shown in Fig.~\ref{fig:DNSP-in-AsNuclear}(a-c). 
Especially when the Overhauser field cancels out the external magnetic
field (i.e. $2A\left\langle I_{z}\right\rangle _{max}\approx g^{*}\mu_{B}B_{ext}$)
in $D_{12}$, as shown for example in Fig.~\ref{fig:zoomin}(b).
At this point the spin pumping rate $D_{12}$ reaches its
maximum, which leads to the peaks around 3 to 4 T in Fig.~\ref{fig:DNSP-in-AsNuclear}.
The peak nuclear spin polarization is
\begin{eqnarray*}
\left\langle I_{z}\right\rangle _{max} & \approx & \frac{1}{2}+\frac{1}{1+\frac{\theta^{2}(\frac{E_{Q}}{E_{z}^{n}})^{2}(\frac{\hbar}{\tau_{c}})^{2}}{\left[(\frac{\hbar}{\tau_{c}})^{2}+(E_{z}^{n})^{2}\right]}}\\
 & \approx & \frac{g^{*}\mu_{B}}{2A}B_{ext}.
\end{eqnarray*}
 Therefore, when $E_{Q}$ increases, the peak polarization $\left\langle I_{z}\right\rangle _{max}$
decreases. The corresponding external field $B_{ext}$ decreases as
well, but that relationship is more complicated because $E_{z}^{n}$
also depends on $B_{ext}$.

Equation~(\ref{eq:ana1}) generally fits well with our numerical
calculations, except for the case when the nuclear Zeeman energy is
equal to the quadrupole energy. When the nuclear Zeeman energy is
nearly resonant with the quadrupole energy, the nuclear spin polarization
is harder to build up because the degeneracy between states 2 and
4 [see Fig. \ref{fig:enrgy-levelEnergy-levels}(b)] causes a maximum
in the depolarization rate $F_{24}$, which is orders of magnitude
larger than all other transition rates. The presence of this large
transition rate equalizes the populations of states 2 and 4: $p_{2}=p_{4}$.
Furthermore, in combination with $F_{12}$, it also prevents the nuclear
spins from accumulating in the highest nuclear spin state. By setting up the master equation of Eq.~(\ref{eq:mastereq}),
we find the steady state solution from the following equations:
\begin{eqnarray*}
F_{12}\: p_{1} & = & (D_{12}+F_{12})\: p_{2},\\
p_{3} & \approx & p_{4},\\
p_{1}+p_{2}+p_{3}+p_{4} & = & 1.
\end{eqnarray*}
 The average NSP is,
\begin{eqnarray}
\left\langle I_{z}\right\rangle  & = & \sum_{i}p_{i}\left\langle I_{i}\right\rangle ,\nonumber \\
 & \approx & \frac{3}{2}\frac{1}{1+4\biggl(\frac{F_{12}}{D_{12}}\biggr)},\label{eq:ana2}
\end{eqnarray}
 where
\begin{eqnarray*}
F_{12} & = & \left(\frac{A_{As}}{N}\right)^{2}\frac{6\tau_{c}f_{e}\theta^{2}}{\hbar^{2}+\tau_{c}^{2}{E_{Q}}^{2}}.
\end{eqnarray*}
 The achievable NSP at the 2-4 degeneracy point depends on the relative
direction of the Overhauser field and the external magnetic field.
When they are parallel, $D_{12}$ is at least one order of magnitude
smaller than $F_{12}$, thus the NSP at this degeneracy is only a
few percent, as shown in Fig.~\ref{fig:DNSP-in-AsNuclear}(d-f).
When the fields are anti-parallel, $D_{12}$ may become comparable
to $F_{12}$, and the resulting NSP strongly depends on the ratio
of $F_{12}$ to $D_{12}$, as shown in Fig.~\ref{fig:DNSP-in-AsNuclear}(a-c) and Fig.~\ref{fig:zoomin}(a).
If this ratio is closer to 1 (such as for QD1 and QD2), the resulting NSP could be as high as 20\%, much higher
than that is achievable in the parallel field case. On the other hand,
if this ratio is far above 1 (in QD3), the resulting NSP is suppressed,
as shown in Fig.~\ref{fig:zoomin}(a), while the field-dependence
becomes abrupt.

\begin{figure*}[t]
\begin{centering}
\subfigure[]{\includegraphics[width=1\columnwidth]{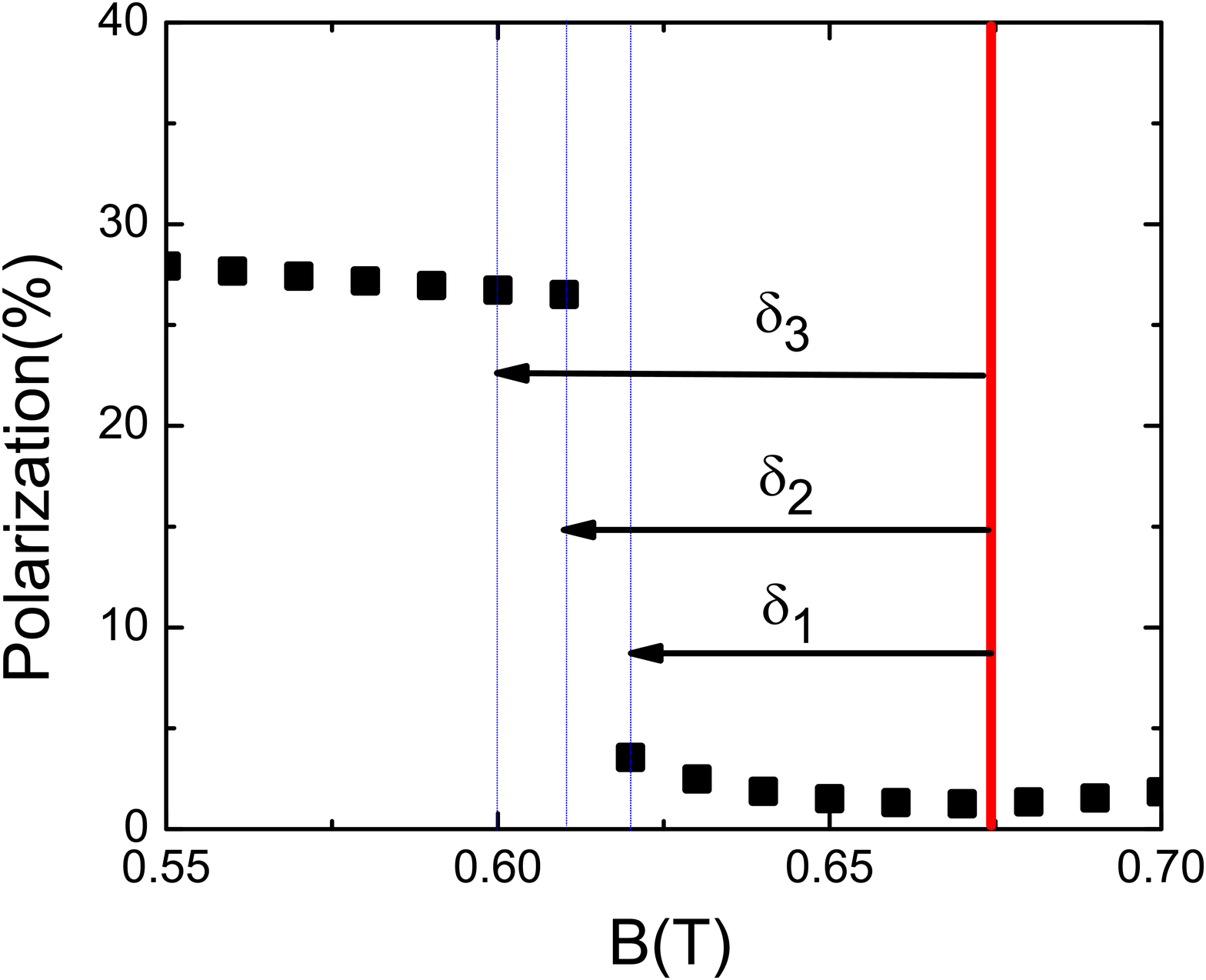}}
\subfigure[]{\includegraphics[width=1\columnwidth]{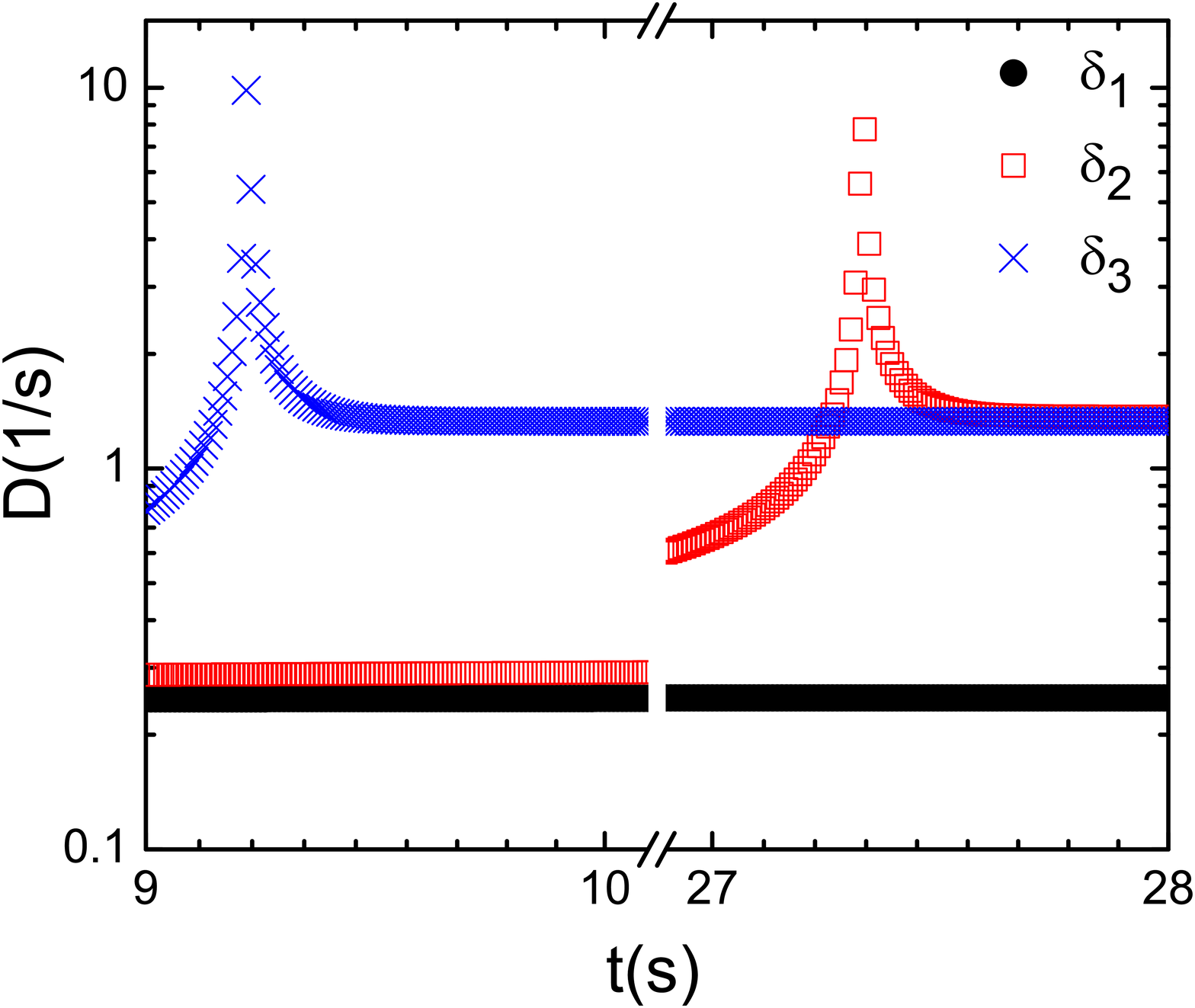}}
\par\end{centering}

\caption{\label{fig:dynamic-changeDynamic-change}(color online). Time evolution of nuclear spin polarization rate [Panel (b)] at different energy detunings [defined
in Panel (a)] in QD3 near the 2-4 degeneracy. The solid line at approximate 0.67 T in Panel (a) stands for the 2-4
degeneracy, where As nuclear Zeeman energy is resonant with the quadrupole energy.  $\delta_{1}$, $\delta_{2}$ and $\delta_{3}$ are 1.7 neV,
1.75 neV, and 1.8 neV from this degeneracy, respectively. Panel (b) shows that NSP build-up time for $\delta_{3}$ is approximately 30 seconds
and 10 seconds for $\delta_{2}$.  For $\delta_{1}$, within our simulation time of 2000 s, the nuclear spin polarization is not built up. Clearly,
as $\delta$ decreases, $F_{24}$ gets closer to its maximum, which is orders of magnitude larger than all other rates, it takes longer and longer
time to build up the NSP, until it cannot be built up.}

\end{figure*}

In order to gain more understanding of the highly nonlinear behavior
of DNSP in QD3, we examine the time evolution of $D_{12}$ at three
different energy detunings, $\delta_{1}$ , $\delta_{2}$ and $\delta_{3}$,
away from the Zeeman-quadrupole resonance, as shown in Fig.~\ref{fig:dynamic-changeDynamic-change}.
According to panel (b), when the pumping rate is high enough to overcome
the depolarization, the Overhauser field starts to build up. When
the Overhauser field cancels out the external magnetic field, the
pumping rate reaches a maximum in the time evolution of the system
[the spikes shown in Fig.~\ref{fig:dynamic-changeDynamic-change}(b)]. 
The Overhauser field quickly exceeds the external magnetic field, 
and then this pumping rate falls off to a steady value, 
in a short time correlated to the hyperfine energy,
cotunneling time constant and $f_{e}$. For $\delta_{3}$, the NSP build-up time is approximately
10 seconds, and for $\delta_{2}$ it is 30 seconds. For $\delta_{1}$,
the pumping never manages to overcome depolarization within our simulation
time (2000 s), and the NSP is limited to a few percent. As the detuning
$\delta$ approaches zero, $F_{24}$ gets closer to its maximum, which
is orders of magnitude larger than all other rates. In this regime,
it takes longer and longer time to build up the nuclear spin polarization,
until it is practically impossible---beyond tens of seconds, nuclear
spin relaxation channels that we do not consider, such as dipolar
induced spin diffusion and direct spin-lattice relaxation, would have
to be included for a complete physical picture to emerge.

When the nuclear Zeeman energy is equal to twice the quadrupole energy,
nuclear spin states 3 and 4 are degenerate, as shown in Fig.~\ref{fig:enrgy-levelEnergy-levels}(c).
However, the NSP around this degeneracy is not as strongly affected
as those near the 2-4 degeneracy. In this field region, $F_{24}$
and $F_{23}$ nearly vanish. The strong $F_{34}$ equalizes the population
of states 3 and 4, so that we can again isolate the four level problem
to a two level one, and NSP can be calculated by using Eq.~\ref{eq:ana1}.

\subsection{Nuclear spin polarization in different compositions of In$_{x}$Ga$_{1-x}$As
quantum dots}

In all the NSP experiments in In$_{x}$Ga$_{1-x}$As QDs, the observable
quantity is the total Overhauser field from all the nuclei of all
the nuclear spin species. Since different nuclear isotopes have
different gyromagnetic ratios (see Table~\ref{tab:elements}), and
different nuclear spins generally experience different electric field
gradients, the DNSP features we study in the previous section for
a single nuclear spin species with a fixed quadrupole splitting would
now occur in ranges of magnetic fields. The total effect is a superposition
of contributions from all the individual ingredients. As we have discussed
in section \ref{sec:Model}, we do not consider interactions between
nuclear spins, whether they are of the same or different species.

\begin{figure*}[t]
\begin{centering}
\includegraphics[width=1.5\columnwidth]{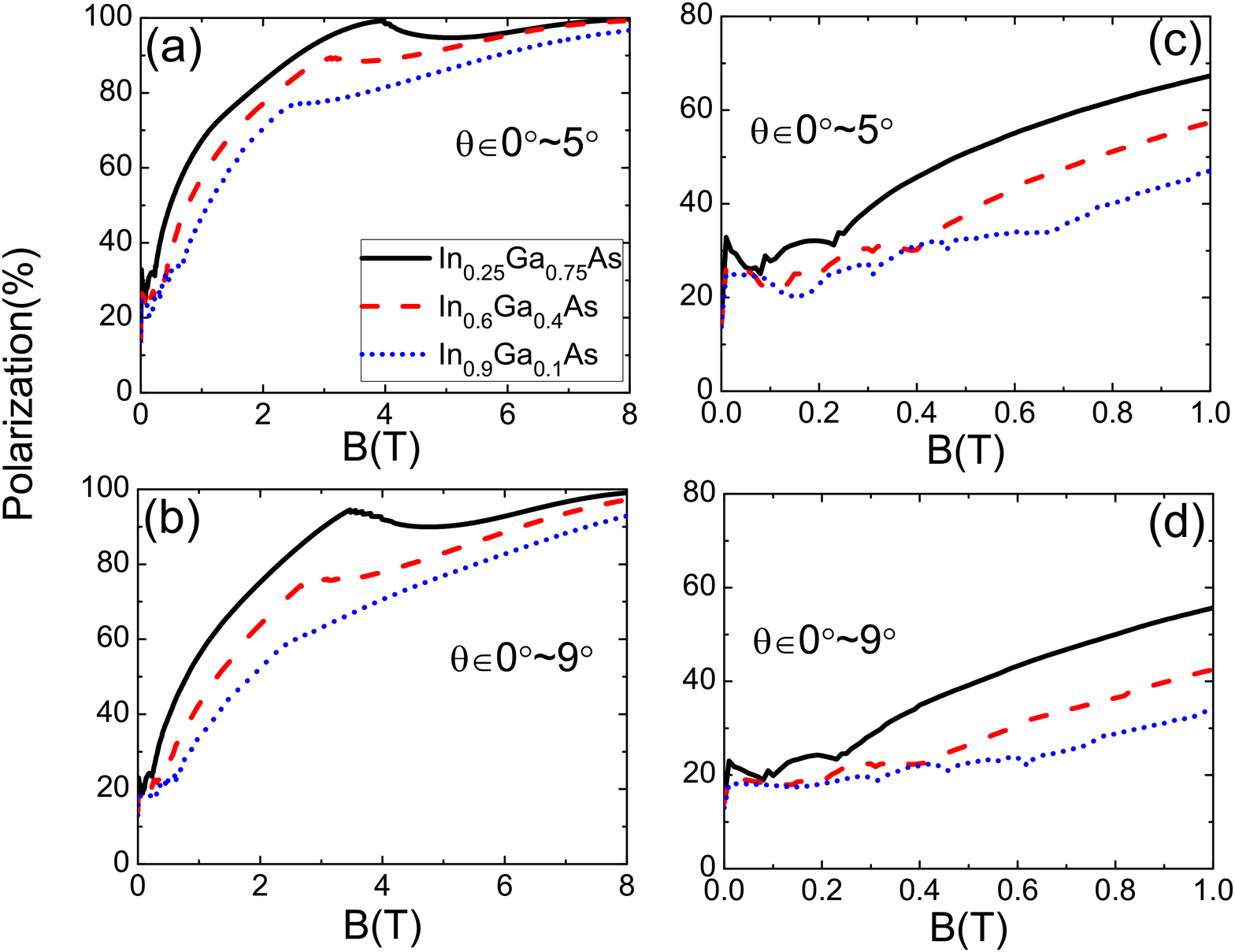}
\par\end{centering}

\caption{\label{fig:InGaAs}(color online). The nuclear spin polarization in In$_{0.25}$Ga$_{0.75}$As (solid lines), 
In$_{0.6}$Ga$_{0.4}$As (dashed lines) and In$_{0.9}$Ga$_{0.1}$As (dotted lines) quantum dots.
 Panels (a) and (c) are for dots with $\theta\in 0-5^{\circ}$ while Panels (b) and (d) are for $\theta\in 0-9^{\circ}$. 
Panels (c) and (d) are zoom-ins to the low field region of Panels (a) and (b), respectively.  
(a) represents the nuclear spin polarization average over angles between $0^{\circ}$ to $5^{\circ}$. 
(b) represents the nuclear spin polarization average over angles between$0^{\circ}$ to $9^{\circ}$. 
(c) and (d) are the detail in low fields for (a) and (b) respectively. 
 Comparing scales of panels (a) with (b), or (c) with (d), we note that stronger strain and larger variations in the
direction of the field gradient suppress the nuclear spin polarization in low to intermediate fields.}

\end{figure*}

To account for the distribution of strain in a QD, we assume a uniform
distribution of angles between the electric field gradient and the
applied magnetic field (which is along $z$ direction as always).
In Fig.~\ref{fig:InGaAs}, the angles between the electric field
gradient and the external magnetic field are in the ranges of $0^{\circ}\sim5^{\circ}$
and $0^{\circ}\sim9^{\circ}$. The peaks and dips in NSP of a single
nuclear spin species, as shown in Fig.~\ref{fig:DNSP-in-AsNuclear},
are now smoothed out, as shown in Fig.~\ref{fig:InGaAs}.

Our results show a qualitative agreement with various experiments.\cite{Eble2006,Maletinsky2007,Tartakovskii2007}
A high degree of nuclear spin polarization can be created in high
field regions, while the polarization is limited in low field regions.
Overall the achievable nuclear spin polarization in In$_{x}$Ga$_{1-x}$As
QDs is related to the concentration of indium and the resulting strain
distribution in the dots. In general, stronger strain and larger angle
between the field gradient and growth direction suppress the nuclear
spin polarization.

\section{\label{sec:Discussion}Discussion}

In our calculations of nuclear spin polarization, the spin pumping is achieved by first optically orienting the electron spins, then
transferring the electron spin polarization to the nuclear spins via either cotunneling-assitsed processes or phonon-assisted processes.  In
the latter we have not included any spin mixing caused by spin-orbit interaction.  However, it is well known that the spin-orbit interaction is
quite strong in InAs dots.\cite{Wingler2003}  Thus we have also explored how the spin-orbit interaction might take part in the
DNSP.\cite{Florescu2006,Khaetskii2000}  More specifically, we have calculated the spin-flip transition rate from the combination of the
hyperfine interaction, spin-orbit interaction, and the electron-phonon interaction, and we find the transition rate is linearly proportional to the
electronic Zeeman splitting. Our results show that the inclusion of spin-orbit interaction into the spin transfer process yields a slower
process (by two orders of magnitude even at the relatively low magnetic field of 1 T) than the hyperfine interaction combined with
electron-phonon interaction alone.  Therefore, we exclude this mechanism from our current study.

\begin{figure}[t]
\begin{centering}
\includegraphics[width=1\columnwidth]{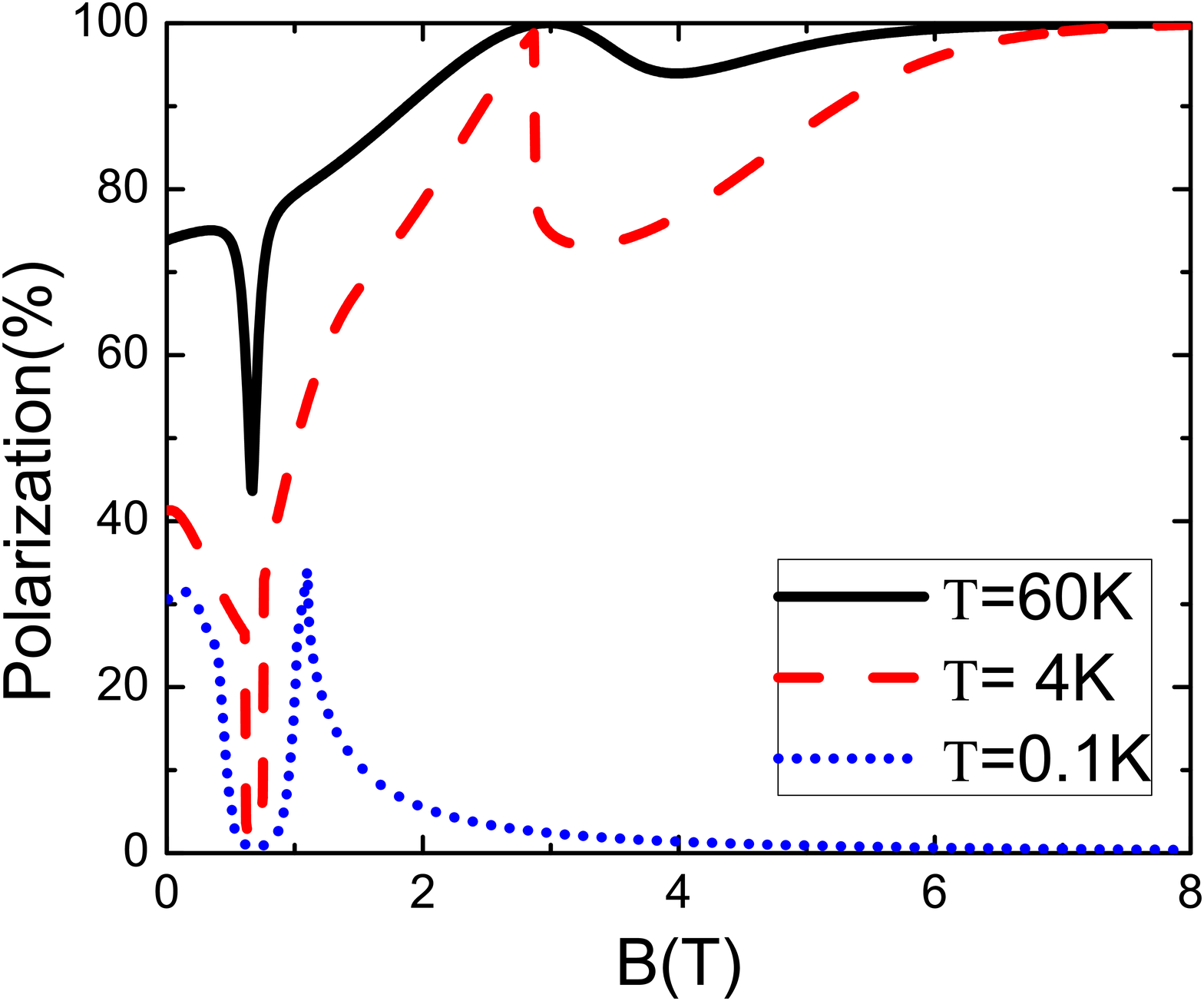}
\par\end{centering}
\caption{\label{fig:tep}(color online). Temperature dependence of As nuclear spin polarization. The solid, dashed and
dotted lines stand for the nuclear spin polarization at temperatures of 0.1 K, 4 K, and 60 K respectively. The nuclear spin polarization is
easier to build up at higher temperature, since the phonon population and cotunneling constants increase as temperature increases.}
\end{figure}

Our calculations presented so far are done at $T = 4$ K.  We have also explored the temperature dependence of the As NSP in a QD.  Both
cotunneling and phonon emission/absorption (especially absorption) are affected by temperature changes, via Fermi level broadening and phonon
populations, so that spin-flip rate will change accordingly.  Consider for example nuclear spins being pumped to the highest-energy state at
high fields, where spin transitions are assisted by phonon absorption, as shown in Fig.~\ref{fig:Hyperfne-induced-transition-among}.  We have
calculated As NSP at three different temperatures: 0.1 K, 4 K and 60 K.  The results are shown in the high field region of Fig.~\ref{fig:tep},
where the NSP can be built up to larger values at temperature 60 K than 4 K and 0.1 K (especially at B > 5 T), since at a fixed magnetic field
the phonon population decreases as temperature decreases ($N\sim\frac{1}{\exp{(E/k_{B}T)}-1}$).  In the low field region, the spin transition is assisted
by cotunneling, and the cotunneling time constant $\tau_c$ is inversely proportional to the temperature.\cite{Smith2005}  Therefore the
cotunneling-assisted processes are more efficient at higher temperatures,\cite{Urbaszek2007} as shown in Fig.~\ref{fig:tep}.
At present we do not have a clear analytical understanding of the abrupt change in NSP as shown in Fig.~\ref{fig:dynamic-changeDynamic-change}.
We are currently working on a full density matrix method that includes all the off-diagonal terms for the electron-nuclear spin system.  Such a
calculation could also help us ascertain the validity of the Master equation approach we adopt in the present study.

\section{\label{sec:Conclusion}Conclusions}

In summary, we have calculated nuclear spin polarization through optical orientation of electron spins in a self-assembled quantum dot. We have
explored how NSP of a single nuclear species depends on the external magnetic field with various strain strengths, angles between the electric
field gradient and the growth direction, and cotunneling energies. We show that, in high magnetic fields, higher degrees of NSP can be
achieved, where the nuclear spin Zeeman energy is much larger than the quadrupole splitting.  In this regime the nuclear spin eigenstates are
close to the eigenstates of $I_{z}$, so that the depolarization rates between nuclear spin states are approximately zero.  Therefore nuclear
spins can be pumped to the highest nuclear energy state without leaking back to lower energy states.  In low to intermediate field regions, NSP
is strongly affected by the strain distribution. Generally speaking, in the same QD, the NSP is lower when the electric field gradient is at a
larger angle from the external magnetic field, because strain along transverse directions (relative to the magnetic field) is the driving force
behind depolarization transitions for the nuclear spins. In addition, NSP is lower at smaller cotunnelling rates (when, for example, the allowed
electronic state in the QD is far below the Fermi sea). Furthermore, NSP is also harder to build up in a QD with a larger magnitude of strain.
Our calculation shows that higher strain strength in a QD leads to smaller NSP in general.

For NSP in In$_{x}$Ga$_{1-x}$As quantum dots, our results are obtained from incoherent superpositions of In, Ga and As contributions in different
proportions.  We show that nearly full nuclear spin polarization can be created in high field regions, while it is limited in low field regions.
Our results indicate that the concentration of indium and the resulting strain distribution in the dots play a crucial role in DNSP.  For
example, at low magnetic fields, nuclear spin polarization is harder to build up in In$_{0.9}$Ga$_{0.1}$As than in In$_{0.6}$Ga$_{0.4}$As.  The
interplay between the nuclear quadrupole interaction and Zeeman splitting could lead to suppression of nuclear spin polarization.  Our results are
in qualitative agreement with the measured nuclear spin polarization in the experimental work of various
groups.\cite{Bracker2005,Eble2006,Maletinsky2008,Maletinsky2007,Tartakovskii2007}

Our results suggest that for a dot with a uniform strain distribution (and with a principal axis away from the external magnetic field), a minimum
in NSP should be expected when the nuclear Zeeman energy is equal to the quadrupole energy.  Moreover, a peak should be observed in the
intermediate field regions (around 3 to 4 T), where the Overhauser field cancels out the external magnetic field.

\begin{acknowledgments}
We thank financial support by NSA/LPS through ARO grants W911NF0610209 and W911NF0910393. XH also acknowledges supports by Joint Quantum
Institute of University of Maryland and DARPA QuEST.
\end{acknowledgments}


\begin{thebibliography}{59}
\expandafter\ifx\csname natexlab\endcsname\relax\def\natexlab#1{#1}\fi
\expandafter\ifx\csname bibnamefont\endcsname\relax
  \def\bibnamefont#1{#1}\fi
\expandafter\ifx\csname bibfnamefont\endcsname\relax
  \def\bibfnamefont#1{#1}\fi
\expandafter\ifx\csname citenamefont\endcsname\relax
  \def\citenamefont#1{#1}\fi
\expandafter\ifx\csname url\endcsname\relax
  \def\url#1{\texttt{#1}}\fi
\expandafter\ifx\csname urlprefix\endcsname\relax\def\urlprefix{URL }\fi
\providecommand{\bibinfo}[2]{#2}
\providecommand{\eprint}[2][]{\url{#2}}

\bibitem[{\citenamefont{Slichter}(1992)}]{Slichter1992}
\bibinfo{author}{\bibfnamefont{C.}~\bibnamefont{Slichter}},
  \emph{\bibinfo{title}{Principles of Magnetic Resonance}}
  (\bibinfo{publisher}{Springer-Verlag}, \bibinfo{year}{1992}).

\bibitem[{\citenamefont{Abragam}(1961)}]{Abragam1961}
\bibinfo{author}{\bibfnamefont{A.}~\bibnamefont{Abragam}},
  \emph{\bibinfo{title}{The Principles of Nuclear Magnetism}}
  (\bibinfo{publisher}{Clarendon Press, Oxford}, \bibinfo{year}{1961}).

\bibitem[{\citenamefont{Kane}(1998)}]{Kane1998}
\bibinfo{author}{\bibfnamefont{B.~E.} \bibnamefont{Kane}},
  \bibinfo{journal}{Nature} \textbf{\bibinfo{volume}{393}},
  \bibinfo{pages}{133} (\bibinfo{year}{1998}).

\bibitem[{\citenamefont{Taylor et~al.}(2003)\citenamefont{Taylor, Marcus, and
  Lukin}}]{Taylor2003}
\bibinfo{author}{\bibfnamefont{J.~M.} \bibnamefont{Taylor}},
  \bibinfo{author}{\bibfnamefont{C.~M.} \bibnamefont{Marcus}},
  \bibnamefont{and} \bibinfo{author}{\bibfnamefont{M.~D.} \bibnamefont{Lukin}},
  \bibinfo{journal}{Phys. Rev. Lett.} \textbf{\bibinfo{volume}{90}},
  \bibinfo{pages}{206803} (\bibinfo{year}{2003}).

\bibitem[{\citenamefont{Loss and DiVincenzo}(1998)}]{Loss1998}
\bibinfo{author}{\bibfnamefont{D.}~\bibnamefont{Loss}} \bibnamefont{and}
  \bibinfo{author}{\bibfnamefont{D.~P.} \bibnamefont{DiVincenzo}},
  \bibinfo{journal}{Phys. Rev. A} \textbf{\bibinfo{volume}{57}},
  \bibinfo{pages}{120} (\bibinfo{year}{1998}).

\bibitem[{\citenamefont{Merkulov et~al.}(2002)\citenamefont{Merkulov, Efros,
  and Rosen}}]{Merkulov2002}
\bibinfo{author}{\bibfnamefont{I.~A.} \bibnamefont{Merkulov}},
  \bibinfo{author}{\bibfnamefont{A.~L.} \bibnamefont{Efros}}, \bibnamefont{and}
  \bibinfo{author}{\bibfnamefont{M.}~\bibnamefont{Rosen}},
  \bibinfo{journal}{Phys. Rev. B} \textbf{\bibinfo{volume}{65}},
  \bibinfo{pages}{205309} (\bibinfo{year}{2002}).

\bibitem[{\citenamefont{Witzel et~al.}(2005)\citenamefont{Witzel, de~Sousa, and
  Das~Sarma}}]{Witzel2005}
\bibinfo{author}{\bibfnamefont{W.~M.} \bibnamefont{Witzel}},
  \bibinfo{author}{\bibfnamefont{R.}~\bibnamefont{de~Sousa}}, \bibnamefont{and}
  \bibinfo{author}{\bibfnamefont{S.}~\bibnamefont{Das~Sarma}},
  \bibinfo{journal}{Phys. Rev. B} \textbf{\bibinfo{volume}{72}},
  \bibinfo{pages}{161306} (\bibinfo{year}{2005}).

\bibitem[{\citenamefont{Coish and Loss}(2004)}]{Coish2004}
\bibinfo{author}{\bibfnamefont{W.~A.} \bibnamefont{Coish}} \bibnamefont{and}
  \bibinfo{author}{\bibfnamefont{D.}~\bibnamefont{Loss}},
  \bibinfo{journal}{Phys. Rev. B} \textbf{\bibinfo{volume}{70}},
  \bibinfo{pages}{195340} (\bibinfo{year}{2004}).

\bibitem[{\citenamefont{Yao et~al.}(2006)\citenamefont{Yao, Liu, and
  Sham}}]{Yao2006}
\bibinfo{author}{\bibfnamefont{W.}~\bibnamefont{Yao}},
  \bibinfo{author}{\bibfnamefont{R.-B.} \bibnamefont{Liu}}, \bibnamefont{and}
  \bibinfo{author}{\bibfnamefont{L.~J.} \bibnamefont{Sham}},
  \bibinfo{journal}{Phys. Rev. B} \textbf{\bibinfo{volume}{74}},
  \bibinfo{pages}{195301} (\bibinfo{year}{2006}).

\bibitem[{\citenamefont{Erlingsson and Nazarov}(2004)}]{Erlingsson2004}
\bibinfo{author}{\bibfnamefont{S.~I.} \bibnamefont{Erlingsson}}
  \bibnamefont{and} \bibinfo{author}{\bibfnamefont{Y.~V.}
  \bibnamefont{Nazarov}}, \bibinfo{journal}{Phys. Rev. B}
  \textbf{\bibinfo{volume}{70}}, \bibinfo{pages}{205327}
  (\bibinfo{year}{2004}).

\bibitem[{\citenamefont{Khaetskii et~al.}(2002)\citenamefont{Khaetskii, Loss,
  and Glazman}}]{Khaetskii2002}
\bibinfo{author}{\bibfnamefont{A.~V.} \bibnamefont{Khaetskii}},
  \bibinfo{author}{\bibfnamefont{D.}~\bibnamefont{Loss}}, \bibnamefont{and}
  \bibinfo{author}{\bibfnamefont{L.}~\bibnamefont{Glazman}},
  \bibinfo{journal}{Phys. Rev. Lett.} \textbf{\bibinfo{volume}{88}},
  \bibinfo{pages}{186802} (\bibinfo{year}{2002}).

\bibitem[{\citenamefont{Burkard et~al.}(1999)\citenamefont{Burkard, Loss,
  DiVincenzo, and Smolin}}]{Burkard1999}
\bibinfo{author}{\bibfnamefont{G.}~\bibnamefont{Burkard}},
  \bibinfo{author}{\bibfnamefont{D.}~\bibnamefont{Loss}},
  \bibinfo{author}{\bibfnamefont{D.~P.} \bibnamefont{DiVincenzo}},
  \bibnamefont{and} \bibinfo{author}{\bibfnamefont{J.~A.}
  \bibnamefont{Smolin}}, \bibinfo{journal}{Phys. Rev. B}
  \textbf{\bibinfo{volume}{60}}, \bibinfo{pages}{11404} (\bibinfo{year}{1999}).

\bibitem[{\citenamefont{Klauser et~al.}(2006)\citenamefont{Klauser, Coish, and
  Loss}}]{Klauser2006}
\bibinfo{author}{\bibfnamefont{D.}~\bibnamefont{Klauser}},
  \bibinfo{author}{\bibfnamefont{W.~A.} \bibnamefont{Coish}}, \bibnamefont{and}
  \bibinfo{author}{\bibfnamefont{D.}~\bibnamefont{Loss}},
  \bibinfo{journal}{Phys. Rev. B} \textbf{\bibinfo{volume}{73}},
  \bibinfo{pages}{205302} (\bibinfo{year}{2006}).

\bibitem[{\citenamefont{Ramon and Hu}(2007)}]{Ramon2007}
\bibinfo{author}{\bibfnamefont{G.}~\bibnamefont{Ramon}} \bibnamefont{and}
  \bibinfo{author}{\bibfnamefont{X.}~\bibnamefont{Hu}}, \bibinfo{journal}{Phys.
  Rev. B} \textbf{\bibinfo{volume}{75}}, \bibinfo{pages}{161301}
  (\bibinfo{year}{2007}).

\bibitem[{\citenamefont{Meier and Azkharchenya}(1984)}]{Meier1984}
\bibinfo{editor}{\bibfnamefont{F.}~\bibnamefont{Meier}} \bibnamefont{and}
  \bibinfo{editor}{\bibfnamefont{B.}~\bibnamefont{Azkharchenya}}, eds.,
  \emph{\bibinfo{title}{Optical Orientation}}
  (\bibinfo{publisher}{North-Holland, Amsterdam}, \bibinfo{year}{1984}).

\bibitem[{\citenamefont{Dobers et~al.}(1988)\citenamefont{Dobers, Klitzing,
  Schneider, Weimann, and Ploog}}]{Dobers1988}
\bibinfo{author}{\bibfnamefont{M.}~\bibnamefont{Dobers}},
  \bibinfo{author}{\bibfnamefont{K.~v.} \bibnamefont{Klitzing}},
  \bibinfo{author}{\bibfnamefont{J.}~\bibnamefont{Schneider}},
  \bibinfo{author}{\bibfnamefont{G.}~\bibnamefont{Weimann}}, \bibnamefont{and}
  \bibinfo{author}{\bibfnamefont{K.}~\bibnamefont{Ploog}},
  \bibinfo{journal}{Phys. Rev. Lett.} \textbf{\bibinfo{volume}{61}},
  \bibinfo{pages}{1650} (\bibinfo{year}{1988}).

\bibitem[{\citenamefont{Kane et~al.}(1992)\citenamefont{Kane, Pfeiffer, and
  West}}]{Kane1992}
\bibinfo{author}{\bibfnamefont{B.~E.} \bibnamefont{Kane}},
  \bibinfo{author}{\bibfnamefont{L.~N.} \bibnamefont{Pfeiffer}},
  \bibnamefont{and} \bibinfo{author}{\bibfnamefont{K.~W.} \bibnamefont{West}},
  \bibinfo{journal}{Phys. Rev. B} \textbf{\bibinfo{volume}{46}},
  \bibinfo{pages}{7264} (\bibinfo{year}{1992}).

\bibitem[{\citenamefont{Smet et~al.}(2002)\citenamefont{Smet, Deutschmann,
  Ertl, Wegscheider, Abstreiter, and von Klitzing}}]{Smet2002}
\bibinfo{author}{\bibfnamefont{J.~H.} \bibnamefont{Smet}},
  \bibinfo{author}{\bibfnamefont{R.~A.} \bibnamefont{Deutschmann}},
  \bibinfo{author}{\bibfnamefont{F.}~\bibnamefont{Ertl}},
  \bibinfo{author}{\bibfnamefont{W.}~\bibnamefont{Wegscheider}},
  \bibinfo{author}{\bibfnamefont{G.}~\bibnamefont{Abstreiter}},
  \bibnamefont{and} \bibinfo{author}{\bibfnamefont{K.}~\bibnamefont{von
  Klitzing}}, \bibinfo{journal}{Nature} \textbf{\bibinfo{volume}{415}},
  \bibinfo{pages}{281} (\bibinfo{year}{2002}).

\bibitem[{\citenamefont{Gammon et~al.}(1996)\citenamefont{Gammon, Snow,
  Shanabrook, Katzer, and Park}}]{Gammon1996}
\bibinfo{author}{\bibfnamefont{D.}~\bibnamefont{Gammon}},
  \bibinfo{author}{\bibfnamefont{E.~S.} \bibnamefont{Snow}},
  \bibinfo{author}{\bibfnamefont{B.~V.} \bibnamefont{Shanabrook}},
  \bibinfo{author}{\bibfnamefont{D.~S.} \bibnamefont{Katzer}},
  \bibnamefont{and} \bibinfo{author}{\bibfnamefont{D.}~\bibnamefont{Park}},
  \bibinfo{journal}{Phys. Rev. Lett.} \textbf{\bibinfo{volume}{76}},
  \bibinfo{pages}{3005} (\bibinfo{year}{1996}).

\bibitem[{\citenamefont{Ono and Tarucha}(2004)}]{Ono2004}
\bibinfo{author}{\bibfnamefont{K.}~\bibnamefont{Ono}} \bibnamefont{and}
  \bibinfo{author}{\bibfnamefont{S.}~\bibnamefont{Tarucha}},
  \bibinfo{journal}{Phys. Rev. Lett.} \textbf{\bibinfo{volume}{92}},
  \bibinfo{pages}{256803} (\bibinfo{year}{2004}).

\bibitem[{\citenamefont{Baugh et~al.}(2007)\citenamefont{Baugh, Kitamura, Ono,
  and Tarucha}}]{Baugh2007}
\bibinfo{author}{\bibfnamefont{J.}~\bibnamefont{Baugh}},
  \bibinfo{author}{\bibfnamefont{Y.}~\bibnamefont{Kitamura}},
  \bibinfo{author}{\bibfnamefont{K.}~\bibnamefont{Ono}}, \bibnamefont{and}
  \bibinfo{author}{\bibfnamefont{S.}~\bibnamefont{Tarucha}},
  \bibinfo{journal}{Phys. Rev. Lett.} \textbf{\bibinfo{volume}{99}},
  \bibinfo{pages}{096804} (\bibinfo{year}{2007}).

\bibitem[{\citenamefont{Lai et~al.}(2006)\citenamefont{Lai, Maletinsky,
  Badolato, and Imamoglu}}]{Lai2006}
\bibinfo{author}{\bibfnamefont{C.~W.} \bibnamefont{Lai}},
  \bibinfo{author}{\bibfnamefont{P.}~\bibnamefont{Maletinsky}},
  \bibinfo{author}{\bibfnamefont{A.}~\bibnamefont{Badolato}}, \bibnamefont{and}
  \bibinfo{author}{\bibfnamefont{A.}~\bibnamefont{Imamoglu}},
  \bibinfo{journal}{Phys. Rev. Lett.} \textbf{\bibinfo{volume}{96}},
  \bibinfo{eid}{167403} (\bibinfo{year}{2006}).

\bibitem[{\citenamefont{Petta et~al.}(2008)\citenamefont{Petta, Taylor,
  Johnson, Yacoby, Lukin, Marcus, Hanson, and Gossard}}]{Petta2008}
\bibinfo{author}{\bibfnamefont{J.~R.} \bibnamefont{Petta}},
  \bibinfo{author}{\bibfnamefont{J.~M.} \bibnamefont{Taylor}},
  \bibinfo{author}{\bibfnamefont{A.~C.} \bibnamefont{Johnson}},
  \bibinfo{author}{\bibfnamefont{A.}~\bibnamefont{Yacoby}},
  \bibinfo{author}{\bibfnamefont{M.~D.} \bibnamefont{Lukin}},
  \bibinfo{author}{\bibfnamefont{C.~M.} \bibnamefont{Marcus}},
  \bibinfo{author}{\bibfnamefont{M.~P.} \bibnamefont{Hanson}},
  \bibnamefont{and} \bibinfo{author}{\bibfnamefont{A.~C.}
  \bibnamefont{Gossard}}, \bibinfo{journal}{Phys. Rev. Lett.}
  \textbf{\bibinfo{volume}{100}}, \bibinfo{pages}{067601}
  (\bibinfo{year}{2008}).

\bibitem[{\citenamefont{Danon et~al.}(2009)\citenamefont{Danon, Vink, Koppens,
  Nowack, Vandersypen, and Nazarov}}]{Danon2009}
\bibinfo{author}{\bibfnamefont{J.}~\bibnamefont{Danon}},
  \bibinfo{author}{\bibfnamefont{I.~T.} \bibnamefont{Vink}},
  \bibinfo{author}{\bibfnamefont{F.~H.~L.} \bibnamefont{Koppens}},
  \bibinfo{author}{\bibfnamefont{K.~C.} \bibnamefont{Nowack}},
  \bibinfo{author}{\bibfnamefont{L.~M.~K.} \bibnamefont{Vandersypen}},
  \bibnamefont{and} \bibinfo{author}{\bibfnamefont{Y.~V.}
  \bibnamefont{Nazarov}}, \bibinfo{journal}{Phys. Rev. Lett.}
  \textbf{\bibinfo{volume}{103}}, \bibinfo{pages}{046601}
  (\bibinfo{year}{2009}).

\bibitem[{\citenamefont{Braun et~al.}(2006)\citenamefont{Braun, Urbaszek,
  Amand, Marie, Krebs, Eble, Lemaitre, and Voisin}}]{Braun2006}
\bibinfo{author}{\bibfnamefont{P.-F.} \bibnamefont{Braun}},
  \bibinfo{author}{\bibfnamefont{B.}~\bibnamefont{Urbaszek}},
  \bibinfo{author}{\bibfnamefont{T.}~\bibnamefont{Amand}},
  \bibinfo{author}{\bibfnamefont{X.}~\bibnamefont{Marie}},
  \bibinfo{author}{\bibfnamefont{O.}~\bibnamefont{Krebs}},
  \bibinfo{author}{\bibfnamefont{B.}~\bibnamefont{Eble}},
  \bibinfo{author}{\bibfnamefont{A.}~\bibnamefont{Lemaitre}}, \bibnamefont{and}
  \bibinfo{author}{\bibfnamefont{P.}~\bibnamefont{Voisin}},
  \bibinfo{journal}{Physical Review B} \textbf{\bibinfo{volume}{74}},
  \bibinfo{eid}{245306} (\bibinfo{year}{2006}).

\bibitem[{\citenamefont{Eble et~al.}(2006)\citenamefont{Eble, Krebs, Lemaitre,
  Kowalik, Kudelski, Voisin, Urbaszek, Marie, and Amand}}]{Eble2006}
\bibinfo{author}{\bibfnamefont{B.}~\bibnamefont{Eble}},
  \bibinfo{author}{\bibfnamefont{O.}~\bibnamefont{Krebs}},
  \bibinfo{author}{\bibfnamefont{A.}~\bibnamefont{Lemaitre}},
  \bibinfo{author}{\bibfnamefont{K.}~\bibnamefont{Kowalik}},
  \bibinfo{author}{\bibfnamefont{A.}~\bibnamefont{Kudelski}},
  \bibinfo{author}{\bibfnamefont{P.}~\bibnamefont{Voisin}},
  \bibinfo{author}{\bibfnamefont{B.}~\bibnamefont{Urbaszek}},
  \bibinfo{author}{\bibfnamefont{X.}~\bibnamefont{Marie}}, \bibnamefont{and}
  \bibinfo{author}{\bibfnamefont{T.}~\bibnamefont{Amand}},
  \bibinfo{journal}{Physical Review B} \textbf{\bibinfo{volume}{74}},
  \bibinfo{eid}{081306} (\bibinfo{year}{2006}).

\bibitem[{\citenamefont{Tartakovskii and et~al.}(2007)}]{Tartakovskii2007}
\bibinfo{author}{\bibfnamefont{A.~I.} \bibnamefont{Tartakovskii}}
  \bibnamefont{and} \bibinfo{author}{\bibnamefont{et~al.}},
  \bibinfo{journal}{Phys. Rev. Lett.} \textbf{\bibinfo{volume}{98}},
  \bibinfo{pages}{026806} (\bibinfo{year}{2007}).

\bibitem[{\citenamefont{Maletinsky}(2008)}]{Maletinsky2008}
\bibinfo{author}{\bibfnamefont{P.}~\bibnamefont{Maletinsky}}, Ph.D. thesis,
  \bibinfo{school}{ETH, Zurich} (\bibinfo{year}{2008}).

\bibitem[{\citenamefont{Dzhioev and Korenev}(2007)}]{Dzhioev2007}
\bibinfo{author}{\bibfnamefont{R.~I.} \bibnamefont{Dzhioev}} \bibnamefont{and}
  \bibinfo{author}{\bibfnamefont{V.~L.} \bibnamefont{Korenev}},
  \bibinfo{journal}{Phys. Rev. Lett.} \textbf{\bibinfo{volume}{99}},
  \bibinfo{pages}{037401} (\bibinfo{year}{2007}).

\bibitem[{\citenamefont{Williamson and Zunger}(1999)}]{Williamson1999}
\bibinfo{author}{\bibfnamefont{A.~J.} \bibnamefont{Williamson}}
  \bibnamefont{and} \bibinfo{author}{\bibfnamefont{A.}~\bibnamefont{Zunger}},
  \bibinfo{journal}{Phys. Rev. B} \textbf{\bibinfo{volume}{59}},
  \bibinfo{pages}{15819} (\bibinfo{year}{1999}).

\bibitem[{\citenamefont{Gammon et~al.}(2001)\citenamefont{Gammon, Efros,
  Kennedy, Rosen, Katzer, Park, Brown, Korenev, and Merkulov}}]{Gammon2001}
\bibinfo{author}{\bibfnamefont{D.}~\bibnamefont{Gammon}},
  \bibinfo{author}{\bibfnamefont{A.~L.} \bibnamefont{Efros}},
  \bibinfo{author}{\bibfnamefont{T.~A.} \bibnamefont{Kennedy}},
  \bibinfo{author}{\bibfnamefont{M.}~\bibnamefont{Rosen}},
  \bibinfo{author}{\bibfnamefont{D.~S.} \bibnamefont{Katzer}},
  \bibinfo{author}{\bibfnamefont{D.}~\bibnamefont{Park}},
  \bibinfo{author}{\bibfnamefont{S.~W.} \bibnamefont{Brown}},
  \bibinfo{author}{\bibfnamefont{V.~L.} \bibnamefont{Korenev}},
  \bibnamefont{and} \bibinfo{author}{\bibfnamefont{I.~A.}
  \bibnamefont{Merkulov}}, \bibinfo{journal}{Phys. Rev. Lett.}
  \textbf{\bibinfo{volume}{86}}, \bibinfo{pages}{5176} (\bibinfo{year}{2001}).

\bibitem[{\citenamefont{Bracker et~al.}(2005)\citenamefont{Bracker, Stinaff,
  Gammon, Ware, Tischler, Shabaev, Efros, Park, Gershoni, Korenev
  et~al.}}]{Bracker2005}
\bibinfo{author}{\bibfnamefont{A.~S.} \bibnamefont{Bracker}},
  \bibinfo{author}{\bibfnamefont{E.~A.} \bibnamefont{Stinaff}},
  \bibinfo{author}{\bibfnamefont{D.}~\bibnamefont{Gammon}},
  \bibinfo{author}{\bibfnamefont{M.~E.} \bibnamefont{Ware}},
  \bibinfo{author}{\bibfnamefont{J.~G.} \bibnamefont{Tischler}},
  \bibinfo{author}{\bibfnamefont{A.}~\bibnamefont{Shabaev}},
  \bibinfo{author}{\bibfnamefont{A.~L.} \bibnamefont{Efros}},
  \bibinfo{author}{\bibfnamefont{D.}~\bibnamefont{Park}},
  \bibinfo{author}{\bibfnamefont{D.}~\bibnamefont{Gershoni}},
  \bibinfo{author}{\bibfnamefont{V.~L.} \bibnamefont{Korenev}},
  \bibnamefont{et~al.}, \bibinfo{journal}{Phys. Rev. Lett.}
  \textbf{\bibinfo{volume}{94}}, \bibinfo{pages}{047402}
  (\bibinfo{year}{2005}).

\bibitem[{\citenamefont{Sundfors}(1974)}]{Sundfors1974}
\bibinfo{author}{\bibfnamefont{R.~K.} \bibnamefont{Sundfors}},
  \bibinfo{journal}{Phys. Rev. B} \textbf{\bibinfo{volume}{10}},
  \bibinfo{pages}{4244} (\bibinfo{year}{1974}).

\bibitem[{\citenamefont{Sundfors et~al.}(1976)\citenamefont{Sundfors, Tsui, and
  Schwab}}]{Sundfors1976}
\bibinfo{author}{\bibfnamefont{R.~K.} \bibnamefont{Sundfors}},
  \bibinfo{author}{\bibfnamefont{R.~K.} \bibnamefont{Tsui}}, \bibnamefont{and}
  \bibinfo{author}{\bibfnamefont{C.}~\bibnamefont{Schwab}},
  \bibinfo{journal}{Phys. Rev. B} \textbf{\bibinfo{volume}{13}},
  \bibinfo{pages}{4504} (\bibinfo{year}{1976}).

\bibitem[{\citenamefont{Weast}(2007)}]{Weast2007}
\bibinfo{author}{\bibfnamefont{R.~C.} \bibnamefont{Weast}},
  \emph{\bibinfo{title}{HandBook of Chemistry and Physics}}
  (\bibinfo{publisher}{The Chemical Rubber Co., Cleveland, OH.},
  \bibinfo{year}{2007}).

\bibitem[{\citenamefont{Grundmann et~al.}(1995)\citenamefont{Grundmann, Stier,
  and Bimberg}}]{Grundmann1995}
\bibinfo{author}{\bibfnamefont{M.}~\bibnamefont{Grundmann}},
  \bibinfo{author}{\bibfnamefont{O.}~\bibnamefont{Stier}}, \bibnamefont{and}
  \bibinfo{author}{\bibfnamefont{D.}~\bibnamefont{Bimberg}},
  \bibinfo{journal}{Phys. Rev. B} \textbf{\bibinfo{volume}{52}},
  \bibinfo{pages}{11969} (\bibinfo{year}{1995}).

\bibitem[{\citenamefont{Korkusinski and Hawrylak}(2001)}]{Korkusinski2001}
\bibinfo{author}{\bibfnamefont{M.}~\bibnamefont{Korkusinski}} \bibnamefont{and}
  \bibinfo{author}{\bibfnamefont{P.}~\bibnamefont{Hawrylak}},
  \bibinfo{journal}{Phys. Rev. B} \textbf{\bibinfo{volume}{63}},
  \bibinfo{pages}{195311} (\bibinfo{year}{2001}).

\bibitem[{\citenamefont{Stier et~al.}(1999)\citenamefont{Stier, Grundmann, and
  Bimberg}}]{Stier1999}
\bibinfo{author}{\bibfnamefont{O.}~\bibnamefont{Stier}},
  \bibinfo{author}{\bibfnamefont{M.}~\bibnamefont{Grundmann}},
  \bibnamefont{and} \bibinfo{author}{\bibfnamefont{D.}~\bibnamefont{Bimberg}},
  \bibinfo{journal}{Phys. Rev. B} \textbf{\bibinfo{volume}{59}},
  \bibinfo{pages}{5688} (\bibinfo{year}{1999}).

\bibitem[{\citenamefont{Sheng and Hawrylak}(2005)}]{Sheng2005}
\bibinfo{author}{\bibfnamefont{W.}~\bibnamefont{Sheng}} \bibnamefont{and}
  \bibinfo{author}{\bibfnamefont{P.}~\bibnamefont{Hawrylak}},
  \bibinfo{journal}{Phys. Rev. B} \textbf{\bibinfo{volume}{72}},
  \bibinfo{pages}{035326} (\bibinfo{year}{2005}).

\bibitem[{\citenamefont{Yang et~al.}(2008)\citenamefont{Yang, Xu, and
  Wang}}]{Yang2008}
\bibinfo{author}{\bibfnamefont{M.}~\bibnamefont{Yang}},
  \bibinfo{author}{\bibfnamefont{S.~J.} \bibnamefont{Xu}}, \bibnamefont{and}
  \bibinfo{author}{\bibfnamefont{J.}~\bibnamefont{Wang}},
  \bibinfo{journal}{Appl. Phys. Lett.} \textbf{\bibinfo{volume}{92}},
  \bibinfo{pages}{083112} (\bibinfo{year}{2008}).

\bibitem[{\citenamefont{Paget et~al.}(2008)\citenamefont{Paget, Amand, and
  Korb}}]{Paget2008}
\bibinfo{author}{\bibfnamefont{D.}~\bibnamefont{Paget}},
  \bibinfo{author}{\bibfnamefont{T.}~\bibnamefont{Amand}}, \bibnamefont{and}
  \bibinfo{author}{\bibfnamefont{J.-P.} \bibnamefont{Korb}},
  \bibinfo{journal}{Physical Review B} \textbf{\bibinfo{volume}{77}},
  \bibinfo{pages}{245201} (\bibinfo{year}{2008}).

\bibitem[{\citenamefont{Deng and Hu}(2005)}]{Deng2005a}
\bibinfo{author}{\bibfnamefont{C.}~\bibnamefont{Deng}} \bibnamefont{and}
  \bibinfo{author}{\bibfnamefont{X.}~\bibnamefont{Hu}}, \bibinfo{journal}{Phys.
  Rev. B} \textbf{\bibinfo{volume}{72}}, \bibinfo{pages}{165333}
  (\bibinfo{year}{2005}).

\bibitem[{\citenamefont{Yusuf and Hu}(2010)}]{Yusuf2010}
\bibinfo{author}{\bibfnamefont{E.}~\bibnamefont{Yusuf}} \bibnamefont{and}
  \bibinfo{author}{\bibfnamefont{X.}~\bibnamefont{Hu}}, \bibinfo{journal}{in
  print}  (\bibinfo{year}{2010}).

\bibitem[{\citenamefont{Deng and Hu}(2006)}]{Deng2006}
\bibinfo{author}{\bibfnamefont{C.}~\bibnamefont{Deng}} \bibnamefont{and}
  \bibinfo{author}{\bibfnamefont{X.}~\bibnamefont{Hu}}, \bibinfo{journal}{Phys.
  Rev. B} \textbf{\bibinfo{volume}{73}}, \bibinfo{pages}{241303}
  (\bibinfo{year}{2006}).

\bibitem[{\citenamefont{Cywinski et~al.}(2009)\citenamefont{Cywinski, Witzel,
  and Das~Sarma}}]{Cywinski2009}
\bibinfo{author}{\bibfnamefont{L.}~\bibnamefont{Cywinski}},
  \bibinfo{author}{\bibfnamefont{W.~M.} \bibnamefont{Witzel}},
  \bibnamefont{and}
  \bibinfo{author}{\bibfnamefont{S.}~\bibnamefont{Das~Sarma}},
  \bibinfo{journal}{Phys. Rev. Lett.} \textbf{\bibinfo{volume}{102}},
  \bibinfo{pages}{057601} (\bibinfo{year}{2009}).

\bibitem[{\citenamefont{Smith et~al.}(2005)\citenamefont{Smith, Dalgarno,
  Warburton, Govorov, Karrai, Gerardot, and Petroff}}]{Smith2005}
\bibinfo{author}{\bibfnamefont{J.~M.} \bibnamefont{Smith}},
  \bibinfo{author}{\bibfnamefont{P.~A.} \bibnamefont{Dalgarno}},
  \bibinfo{author}{\bibfnamefont{R.~J.} \bibnamefont{Warburton}},
  \bibinfo{author}{\bibfnamefont{A.~O.} \bibnamefont{Govorov}},
  \bibinfo{author}{\bibfnamefont{K.}~\bibnamefont{Karrai}},
  \bibinfo{author}{\bibfnamefont{B.~D.} \bibnamefont{Gerardot}},
  \bibnamefont{and} \bibinfo{author}{\bibfnamefont{P.~M.}
  \bibnamefont{Petroff}}, \bibinfo{journal}{Phys. Rev. Lett.}
  \textbf{\bibinfo{volume}{94}}, \bibinfo{pages}{197402}
  (\bibinfo{year}{2005}).

\bibitem[{\citenamefont{Schrieffer and Wolff}(1966)}]{Schrieffer1966}
\bibinfo{author}{\bibfnamefont{J.~R.} \bibnamefont{Schrieffer}}
  \bibnamefont{and} \bibinfo{author}{\bibfnamefont{P.~A.} \bibnamefont{Wolff}},
  \bibinfo{journal}{Phys. Rev.} \textbf{\bibinfo{volume}{149}},
  \bibinfo{pages}{491} (\bibinfo{year}{1966}).

\bibitem[{\citenamefont{Mahan}(2000)}]{Mahan2000}
\bibinfo{author}{\bibfnamefont{G.}~\bibnamefont{Mahan}},
  \emph{\bibinfo{title}{Many Particle Physics}} (\bibinfo{publisher}{Kluwer
  Academic/Plenum Publishers, New York}, \bibinfo{year}{2000}).

\bibitem[{\citenamefont{Maletinsky et~al.}(2007)\citenamefont{Maletinsky,
  Badolato, and Imamoglu}}]{Maletinsky2007}
\bibinfo{author}{\bibfnamefont{P.}~\bibnamefont{Maletinsky}},
  \bibinfo{author}{\bibfnamefont{A.}~\bibnamefont{Badolato}}, \bibnamefont{and}
  \bibinfo{author}{\bibfnamefont{A.}~\bibnamefont{Imamoglu}},
  \bibinfo{journal}{Phys. Rev. Lett.} \textbf{\bibinfo{volume}{99}},
  \bibinfo{pages}{056804} (\bibinfo{year}{2007}).

\bibitem[{\citenamefont{Warburton et~al.}(2000)\citenamefont{Warburton,
  Schaflein, Haft, Bickel, Lorke, Karrai, Garcia, Schoenfeld, and
  Petroff}}]{Warburton2000}
\bibinfo{author}{\bibfnamefont{R.~J.} \bibnamefont{Warburton}},
  \bibinfo{author}{\bibfnamefont{C.}~\bibnamefont{Schaflein}},
  \bibinfo{author}{\bibfnamefont{D.}~\bibnamefont{Haft}},
  \bibinfo{author}{\bibfnamefont{F.}~\bibnamefont{Bickel}},
  \bibinfo{author}{\bibfnamefont{A.}~\bibnamefont{Lorke}},
  \bibinfo{author}{\bibfnamefont{K.}~\bibnamefont{Karrai}},
  \bibinfo{author}{\bibfnamefont{J.~M.} \bibnamefont{Garcia}},
  \bibinfo{author}{\bibfnamefont{W.}~\bibnamefont{Schoenfeld}},
  \bibnamefont{and} \bibinfo{author}{\bibfnamefont{P.~M.}
  \bibnamefont{Petroff}}, \bibinfo{journal}{Nature}
  \textbf{\bibinfo{volume}{405}}, \bibinfo{pages}{926} (\bibinfo{year}{2000}).

\bibitem[{\citenamefont{Dreiser et~al.}(2008)\citenamefont{Dreiser, Atature,
  Galland, Muller, Badolato, and Imamoglu}}]{Dreiser2008}
\bibinfo{author}{\bibfnamefont{J.}~\bibnamefont{Dreiser}},
  \bibinfo{author}{\bibfnamefont{M.}~\bibnamefont{Atature}},
  \bibinfo{author}{\bibfnamefont{C.}~\bibnamefont{Galland}},
  \bibinfo{author}{\bibfnamefont{T.}~\bibnamefont{Muller}},
  \bibinfo{author}{\bibfnamefont{A.}~\bibnamefont{Badolato}}, \bibnamefont{and}
  \bibinfo{author}{\bibfnamefont{A.}~\bibnamefont{Imamoglu}},
  \bibinfo{journal}{Physical Review B} \textbf{\bibinfo{volume}{77}},
  \bibinfo{eid}{075317} (\bibinfo{year}{2008}).

\bibitem[{\citenamefont{Erlingsson and Nazarov}(2002)}]{Erlingsson2002}
\bibinfo{author}{\bibfnamefont{S.~I.} \bibnamefont{Erlingsson}}
  \bibnamefont{and} \bibinfo{author}{\bibfnamefont{Y.~V.}
  \bibnamefont{Nazarov}}, \bibinfo{journal}{Phys. Rev. B}
  \textbf{\bibinfo{volume}{66}}, \bibinfo{pages}{155327}
  (\bibinfo{year}{2002}).

\bibitem[{\citenamefont{Migliorato et~al.}(2002)\citenamefont{Migliorato,
  Cullis, Fearn, and Jefferson}}]{Migliorato2002}
\bibinfo{author}{\bibfnamefont{M.~A.} \bibnamefont{Migliorato}},
  \bibinfo{author}{\bibfnamefont{A.~G.} \bibnamefont{Cullis}},
  \bibinfo{author}{\bibfnamefont{M.}~\bibnamefont{Fearn}}, \bibnamefont{and}
  \bibinfo{author}{\bibfnamefont{J.~H.} \bibnamefont{Jefferson}},
  \bibinfo{journal}{Phys. Rev. B} \textbf{\bibinfo{volume}{65}},
  \bibinfo{pages}{115316} (\bibinfo{year}{2002}).

\bibitem[{\citenamefont{Pryor and Flatte}(2006)}]{Pryor2006}
\bibinfo{author}{\bibfnamefont{C.~E.} \bibnamefont{Pryor}} \bibnamefont{and}
  \bibinfo{author}{\bibfnamefont{M.~E.} \bibnamefont{Flatte}},
  \bibinfo{journal}{Phys. Rev. Lett.} \textbf{\bibinfo{volume}{96}},
  \bibinfo{eid}{026804} (\bibinfo{year}{2006}).

\bibitem[{\citenamefont{Bayer et~al.}(2002)\citenamefont{Bayer, Ortner, Stern,
  Kuther, Gorbunov, Forchel, Hawrylak, Fafard, Hinzer, Reinecke
  et~al.}}]{Bayer2002}
\bibinfo{author}{\bibfnamefont{M.}~\bibnamefont{Bayer}},
  \bibinfo{author}{\bibfnamefont{G.}~\bibnamefont{Ortner}},
  \bibinfo{author}{\bibfnamefont{O.}~\bibnamefont{Stern}},
  \bibinfo{author}{\bibfnamefont{A.}~\bibnamefont{Kuther}},
  \bibinfo{author}{\bibfnamefont{A.~A.} \bibnamefont{Gorbunov}},
  \bibinfo{author}{\bibfnamefont{A.}~\bibnamefont{Forchel}},
  \bibinfo{author}{\bibfnamefont{P.}~\bibnamefont{Hawrylak}},
  \bibinfo{author}{\bibfnamefont{S.}~\bibnamefont{Fafard}},
  \bibinfo{author}{\bibfnamefont{K.}~\bibnamefont{Hinzer}},
  \bibinfo{author}{\bibfnamefont{T.~L.} \bibnamefont{Reinecke}},
  \bibnamefont{et~al.}, \bibinfo{journal}{Phys. Rev. B}
  \textbf{\bibinfo{volume}{65}}, \bibinfo{pages}{195315}
  (\bibinfo{year}{2002}).

\bibitem[{\citenamefont{Wingler}(2003)}]{Wingler2003}
\bibinfo{author}{\bibfnamefont{R.}~\bibnamefont{Wingler}},
  \emph{\bibinfo{title}{Spin--Orbit Coupling Effects in Two-Dimensional
  Electron and Hole Systems}} (\bibinfo{publisher}{Springer Berlin /
  Heidelberg}, \bibinfo{year}{2003}).

\bibitem[{\citenamefont{Florescu and Hawrylak}(2006)}]{Florescu2006}
\bibinfo{author}{\bibfnamefont{M.}~\bibnamefont{Florescu}} \bibnamefont{and}
  \bibinfo{author}{\bibfnamefont{P.}~\bibnamefont{Hawrylak}},
  \bibinfo{journal}{Physical Review B} \textbf{\bibinfo{volume}{73}},
  \bibinfo{eid}{045304} (\bibinfo{year}{2006}).

\bibitem[{\citenamefont{Khaetskii and Nazarov}(2000)}]{Khaetskii2000}
\bibinfo{author}{\bibfnamefont{A.~V.} \bibnamefont{Khaetskii}}
  \bibnamefont{and} \bibinfo{author}{\bibfnamefont{Y.~V.}
  \bibnamefont{Nazarov}}, \bibinfo{journal}{Phys. Rev. B}
  \textbf{\bibinfo{volume}{61}}, \bibinfo{pages}{12639} (\bibinfo{year}{2000}).

\bibitem[{\citenamefont{Urbaszek et~al.}(2007)\citenamefont{Urbaszek, Braun,
  Amand, Krebs, Belhadj, Lemaítre, Voisin, and Marie}}]{Urbaszek2007}
\bibinfo{author}{\bibfnamefont{B.}~\bibnamefont{Urbaszek}},
  \bibinfo{author}{\bibfnamefont{P.-F.} \bibnamefont{Braun}},
  \bibinfo{author}{\bibfnamefont{T.}~\bibnamefont{Amand}},
  \bibinfo{author}{\bibfnamefont{O.}~\bibnamefont{Krebs}},
  \bibinfo{author}{\bibfnamefont{T.}~\bibnamefont{Belhadj}},
  \bibinfo{author}{\bibfnamefont{A.}~\bibnamefont{Lemaítre}},
  \bibinfo{author}{\bibfnamefont{P.}~\bibnamefont{Voisin}}, \bibnamefont{and}
  \bibinfo{author}{\bibfnamefont{X.}~\bibnamefont{Marie}},
  \bibinfo{journal}{Phys. Rev. B} \textbf{\bibinfo{volume}{76}},
  \bibinfo{pages}{201301} (\bibinfo{year}{2007}).

\end{thebibliography}
\end{document}